%% file: aph.tex
\documentclass[iop]{emulateapj}
\usepackage{natbib}
\usepackage{comment}
\usepackage{amsmath}
\usepackage{graphicx}

\shorttitle{Sizes of $\mu$Jy Radio Sources}
\shortauthors{MURPHY ET AL.}
\slugcomment{Accepted Version (rev2); \today}

\begin{document}
%\title{The GOODS-N JVLA 10\,GHz Survey: I. Sizes of $\mu$Jy Radio Sources}
\title{The GOODS-N Jansky VLA 10\,GHz Pilot Survey: Sizes of Star-Forming $\mu$Jy Radio Sources}

\author{Eric J. Murphy\altaffilmark{1,2}, Emmanuel Momjian\altaffilmark{3}, James J. Condon\altaffilmark{1}, Ranga-Ram Chary\altaffilmark{2}, Mark Dickinson\altaffilmark{4}, Hanae Inami\altaffilmark{4}, Andrew R. Taylor\altaffilmark{5,6}, and Benjamin J. Weiner\altaffilmark{7}}

\altaffiltext{1}{National Radio Astronomy Observatory, 520 Edgemont Road, Charlottesville, VA 22903, USA; \email{emurphy@nrao.edu}}
\altaffiltext{2}{Infrared Processing and Analysis Center, California Institute of Technology, MC 314-6, Pasadena, CA 91125, USA}
\altaffiltext{3}{National Radio Astronomy Observatory, P.O. Box O, 1003 Lopezville Road, Socorro, NM 87801, USA}
\altaffiltext{4}{National Optical Astronomy Observatories, 950 N Cherry Avenue, Tucson, AZ 85719, USA}
\altaffiltext{5}{Department of Astronomy, University of Cape Town, Rondebosch 7005, Republic of South Africa}
\altaffiltext{6}{Department of Physics, University of the Western Cape, Belleville 7535, Republic of South Africa}
\altaffiltext{7}{Steward Observatory, Department of Astronomy, University of Arizona, AZ 85721, USA}

\begin{abstract}
Our sensitive ($\sigma_\mathrm{n}\approx572\mathrm{\,nJy\,beam}^{-1}$), high-resolution (FWHM~$\theta_{1/2}=0\,\farcs22\approx2\mathrm{\,kpc~at~}z\gtrsim1$) 10\,GHz image covering a single Karl G.~Jansky Very Large Array (VLA) primary beam (FWHM~$\Theta_{1/2}\approx 4\farcm25$) in the GOODS-N field contains 32 sources with $S_\mathrm{p}\gtrsim2\,\mu\mathrm{Jy~beam}^{-1}$ and optical and/or near-infrared (OIR) counterparts.  
Most are about as large as the star-forming regions that power them.  
Their median FWHM major axis is $\langle\theta_\mathrm{M} \rangle=167\pm32 \mathrm{\,mas} \approx1.2\pm0.28\mathrm{\,kpc}$ with rms scatter $\approx91\mathrm{\,mas} \approx 0.79 \mathrm{\,kpc}$.  
In units of the effective radius $r_{\rm e}$ that encloses half their flux, these radio sizes are $\langle r_{\rm e}\rangle\approx 69\pm13\mathrm{\,mas}\approx509\pm114\mathrm{\,pc}$ and have rms scatter $\approx38\mathrm{\,mas}\approx324\mathrm{\,pc}$. 
These sizes are smaller than those measured at lower radio frequencies, but agree with dust emission sizes measured at mm/sub-mm wavelengths and extinction-corrected H$\alpha$ sizes.  
We made a low-resolution ($\theta_{1/2}=1\farcs0$) image with $\approx10\times$ better brightness sensitivity to detect extended sources and measure matched-resolution spectral indices $\alpha_{\rm1.4\,GHz}^{\rm10\,GHz}$.  
It contains 6 new sources with $S_\mathrm{p}\gtrsim3.9\,\mu\mathrm{Jy~beam}^{-1}$ and OIR counterparts.  
The median redshift of all 38 sources is $\langle z\rangle=1.24\pm0.15$.  
The 19 sources with 1.4\,GHz counterparts have median spectral index $\langle\alpha_{\rm 1.4\,GHz}^{\rm10\,GHz}\rangle=-0.74\pm0.10$ with rms scatter $\approx0.35$.  
Including upper limits on $\alpha$ for sources not detected at 1.4\,GHz flattens the median to $\langle\alpha_{\rm 1.4\,GHz}^{\rm10\,GHz}\rangle\gtrsim-0.61$, suggesting that the $\mu$Jy radio sources at higher redshifts, and hence selected at higher rest-frame frequencies, may have flatter spectra.  
If the non-thermal spectral index is $\alpha_{\rm NT} \approx -0.85$, the median thermal fraction of sources selected at median rest-frame frequency $\approx20$\,GHz is $\gtrsim$48\%.

\end{abstract}

\section{Introduction}
Most radio surveys aimed at measuring the star-formation history of
the universe have been made at 1.4 or 3\,GHz
%%REF1
% {\bf  \citep[e.g.,][]{es07, gm10, vsmol16}, }
 \citealp[(e.g.,][]{es07, gm10}; Smol{\v c}i{\'c} et al. 2016, in press), 
 so they are more sensitive to steep-spectrum synchrotron radiation than flat-spectrum free-free
emission.  
Low frequencies were favored because
telescope primary beam areas decrease with 
%XX JJC: deleted as redundant: increasing 
frequency
%XX JJC: Eric: I revised this paragraph per your ``please respond''
%XX to the referee's comment about the survey speed discussion being 
%XX ``a bit bold'', by which I think he means ``too brief''.
($\Omega_{\rm PB} \propto \nu^{-2}$) and the steep decimetric spectra
of star-forming galaxies $\langle \alpha \rangle \approx -0.7$, where
$S_\nu \propto \nu^\alpha$ \citep{jc92}, often cause ``survey speed'' to decline sharply
with frequency:  $SS \propto \nu^{-3.4}$ if system temperature, bandwidth, etc. 
are fixed.  For a detailed quantitative discussion, see \citet{jc-vlass}.
%XX JJC: Eric: put the condon2015 reference into the bibliography:
%Condon, J. 2015, arXiv150.205616
Although more difficult to detect, higher-frequency radio emission from galaxies provides
independent information on galaxy energetics and the star-formation
process.  Constructing the star-formation history of the universe
requires converting the synchrotron luminosities measured by low
frequency surveys to star-formation rates via the tight, but empirical
and local,
%XX JJC: inserted comma above
far-infrared/radio correlation \citep{de85,gxh85,yrc01}.
Surveys at observing frequencies $\gtrsim$10\,GHz 
%XX JJC: I moved ``yield higher angular resolution,'' later in the sentence
%XX JJC: to avoid breaking up your physical argument
measure flux densities closer to the rest-frame
frequencies $\nu \gtrsim 30 \mathrm{\,GHz}$, where the total radio
emission is dominated by the free-free radiation that is more directly
proportional to the rate of massive star formation
\citep[e.g.,][]{pgm67a,kg86,th83,th85,kj99,ejm12b,nb12,ejm15}, 
are still unbiased by dust emission or absorption,
and yield higher angular resolution
%XX JJC: inserted to satisfy referee comment
for a given array size.

%Although the %pre-upgraded 
%XX JJCed: replaced ``pre-upgraded'' by
%original 
%VLA was less sensitive, 
Using the original VLA, \citet{richards98,richards99} 
reached $\sigma_\mathrm{n} = 8.5\,\mu
\mathrm{Jy~beam}^{-1}$ at 8.5\,GHz in the Hubble Deep Field-North
\citep[HDF-N,][]{hdf96}. 
%The significantly increased bandwidth delivered by the Wideband Interferometric Digital ARchitecture (WIDAR) correlator on the upgraded VLA   
The significantly increased bandwidth 
%XX JJCed: deleted excess technical detail ``delivered by the wide-band receivers and the electronics 
of the upgraded VLA
%and processed by its Wideband Interferometric Digital ARchitecture (WIDAR) correlator, 
now allows more sensitive surveys 
%at frequencies up to $\nu \sim 10$\,GHz 
in a reasonable amount of observing time.
Furthermore, the VLA will remain the only radio interferometer able to conduct such high-frequency radio continuum surveys until the SKA1-MID comes online equipped with the Band-5a/b ($5-9.25$\,GHz/$9-16.7$\,GHz)
%Band-5 ($4.6-13.8$\,GHz) 
receivers perhaps in $\gtrsim$2025 \citep[e.g., ][]{ejm_skabk}.

In this paper, we present initial results 
%%REF1
on a flux-limited sample of galaxies from pilot observations aimed at mapping the entire Great 
Observatories Origins Deep Survey-North \citep[GOODS-N;][]{goods03,mg04} field
at 10\,GHz.  The GOODS-N field at J2000 $\alpha
=12^\mathrm{h}36^\mathrm{m}55^\mathrm{s}, \delta =
+62\degr14\arcmin15\arcsec$ covers $\approx 160$\,arcmin$^{2}$ centered
on the HDF-N and is unrivaled in terms of its ancillary data.  These
include extremely deep {\it Chandra}, {\it Hubble Space Telescope}
({\it HST}), {\it Spitzer}, and {\it Herschel} observations, deep
$UBVRIJHK$ ground-based imaging, $\sim$3500 spectroscopic redshifts
from 8--10\,m telescopes, and some of the deepest 1.4\,GHz
observations ever made \citep{gm10}.  Our new VLA X-band ($8-12$\,GHz; reference frequency 10\,GHz) 
%10\,GHz
pilot image has $\approx 15 \times$ better point-source sensitivity
than the 
%XX JJCed: deleted as redundat: existing 
\citet{richards98} image.  
%XX JJCed: shortened: Additionally, with $\theta_{1/2} \approx 0\,\farcs22$ angular resolution, it 
Its $\theta_{1/2} \approx 0\,\farcs22$ angular resolution
is well matched to 
the resolution of 
%XX JJC: deleted as redundant: existing 
{\it HST}/ACS optical and {\it HST}/WFC3-IR
(continuum + H$\alpha$ imaging) data from GOODS and the Cosmic Assembly Near-Infrared Deep Extragalactic Legacy Survey \citep[CANDELS;][]{candels11,candels11_imgs}, and
delivers a physical resolution of $\lesssim 1.9 \mathrm{\,kpc}$ at
{\it any} redshift.  Finally, the radio data provide an
extinction-free view of the morphologies of dusty 
%XX JJCed: replaced ``star-bursting'' by
starburst
galaxies 
%XX JJCed: replaced ``that'' by
which dominate the cosmic star-formation activity between $1 \lesssim
z \lesssim 3$ \citep[e.g.,][]{ejm11a,bm13}.  In this redshift range, 
10\,GHz observations sample
%XX JJC: replaced ``$10 \lesssim \nu \lesssim 40 \mathrm{\,GHz}$'' by
$20 \lesssim \nu \lesssim 40 \mathrm{\,GHz}$ 
in the source frame, where galaxy
emission is expected to be dominated by free-free radiation
\citep[e.g.,][]{ejm09c} and thus provide accurate star-formation
rates for comparison with other diagnostics available from the GOODS
ancillary data (e.g., FUV continuum, H$\alpha$, [O{\sc iii]}5007\,\AA).

In this paper we highlight our findings on the typical 10\,GHz source
characteristics based on these new, extremely deep VLA data.
The paper is organized as follows: In \S2 we describe the data as well
as the analysis procedures used in the present study.  In \S3 we
present our results and discuss their implications.  Our main
conclusions are then summarized in \S4.  All calculations are made
assuming a Hubble constant $H_0 = 71 \mathrm{~km~s}^{-1}
\mathrm{~Mpc}^{-1}$
%XX JJC: deleted precedingcomma 
and a 
%XX JJC: replaced ``standard'' by the more descriptive
flat 
$\Lambda$CDM cosmology with
$\Omega_{\rm M} = 0.27$ and $\Omega_{\Lambda} = 0.73$.

\section{Data and Analysis}

In this section we describe our observations and imaging procedure.
We additionally describe in detail our source-finding and optical/NIR (OIR) 
cross-matching procedure used to create a highly reliable final sample
of sources.

\subsection{Observations}
%XX JJC: removed 'in the GOOOD-N field' to avoid implying the GOODS-N field
%XX is centered on our pointing position
We observed a single pointing centered on J2000 $\alpha=12^\mathrm{h}36^\mathrm{m}51\fs26, \delta = +62\degr13\arcmin37\farcs4$ over the $8-12$\,GHz X-band frequency range 
%XX JJC: replaced 'in' by 'with':
with
the A- and C-configurations as part of the project VLA/14B-037.  
We chose this pointing center to maximize the overlap of known sources in GOODS-N detected at other frequencies, 
%XX JJC: replaced 'specifically continuum detections from a program that carried out blind redshifted CO ($J=1\rightarrow0$) observations with the VLA (project VLA/13A-398; PI: Riechers) in the $31-39$\,GHz frequency range.' by 
specifically the $31-39$\,GHz
continuum detections from VLA project VLA/13A-398 (PI: Riechers)
that searched for redshifted 115~GHz CO $J=1\rightarrow0$.  
%XX JJC: shortened sentence below
These  continuum detection results (J. Hodge et al., in preparation) remain unpublished and are thus not included in the present analysis.
We utilized two pairs of the 3-bit samplers of the VLA, each with 2\,GHz bandwidth and dual polarization (R and L). 
For each sampler pair the Wideband Interferometric Digital ARchitecture (WIDAR) correlator delivered 16 sub-bands, each 128\,MHz wide with 2\,MHz spectral channels and full polarization products (RR, LL, RL, LR).

The on-source integration times in the A and C configurations were
roughly 23 and 1.5\,hr, respectively.  The A-configuration
observations were carried out over seven separate runs during 2015
June, and the C-configuration observations were made during a single
run in 2014 December.  The source 3C\,286 was used as the flux density scale and
bandpass calibrator, while J1302+5748 was used as the complex gain and
telescope pointing calibrator.  Full polarization information was
additionally obtained, using 3C\,286 to calibrate the polarization
position angle and J1407+2827 as a instrumental polarization (leakage) calibrator.  However, the
polarization results will be deferred to a future paper.  
To reduce these data, we used the Common Astronomy Software Applications \citep[CASA;][]{casa} package and followed standard calibration and editing procedures, including the utilization of the VLA calibration pipeline.  

%XX JJC: This section doesn't clearly state that the A-configuration
%XX images include C-array data to fill in the hole in the
%XX (u,v) plane.  Similarly, Section 3 (Results and Discussion) says only that
%XX we used the A configuration. Did we include C-array spacings in the A-array image?
%XX I hope so.  Otherwise the large hole in the (u,v) palne
%XX will reduce the integrated flux densities of extended sources in the 1'' and,
%XX 2'' images, even after tapering, multiscale clean, etc.  
\subsection{Imaging \label{sec:imging}}
The calibrated A- and C-configuration measurement sets were imaged together using the task {\sc tclean} in CASA version 4.6.  
The inclusion of the C-configuration data helps to fill in the hole in the $(u,v)$-plane left by the A-configuration data alone that, if not accounted for, will reduce the integrated flux densities of extended sources. 
The mode of {\sc tclean} was set to multifrequency synthesis \citep[{\sc mfs};][]{mfs1,mfs2}.  
After significant experimentation, we chose to use Briggs weighting with {\sc robust=0.5} and set the parameter {\sc nterms=2}.  
{\sc nterms} is the number of Taylor terms to model the frequency dependence of the sky emission.
The value of 2 allows the {\sc mfs} cleaning procedure to fit sources with different spectral indices in addition to the sources' intensities.
The use of Briggs weighting is necessary to suppress the broad pedestal of the
naturally-weighted dirty beam of the VLA A configuration, and thus
helps to suppress sidelobes.  To deconvolve extended low-intensity emission, we took advantage of the multiscale clean option \citep{msclean,msmfs} in CASA, searching for structures with 
scales up to $\approx 16 \times$ the FWHM of the synthesized beam (i.e., about $\approx 2 \times$ larger than the FWHM of the synthesized beam in the C configuration).  
We additionally invoked the W-projection algorithm \citep{wproJ1,wproj2} using 16 W-planes to take the non-coplanar nature of the array into account.

To improve the quality of the dirty beam by both making its shape more
nearly Gaussian and further suppressing sidelobe structure, we applied
a taper for baselines longer than $\sim 10^6 \lambda$.  This taper was
found to produce the best combination of brightness-temperature and point-source sensitivity.  
The main lobe of the dirty beam was nearly a circular Gaussian 
(major and minor FWHMs $\theta_\mathrm{maj} = 0\,\farcs223 \times \theta_\mathrm{min} = 0\,\farcs206$); 
thus, for simplicity, we restored the final image with a circular Gaussian with FWHM $\theta_{1/2} = 0\,\farcs22$.  
The final high-resolution image used in this analysis is a $10\arcmin$ 
on a side square and within 5\% of the primary beam response has an rms noise $\sigma_\mathrm{n} \approx 572 \mathrm{\,nJy\,beam}^{-1} \approx 139 \mathrm{\,mK}$ 
%%REF1
at the image center.  
For a primary beam FWHM at 10\,GHz of $\Theta_{1/2} \approx 4 \farcm 25$, the image is approximately $2.35 \times \Theta_{1/2}$ on a side.  
% (i.e., $\approx 2.35 \times \Theta_{1/2}$, where $\Theta_{1/2} \approx 4 \farcm 25$ is the FWHM of the primary beam at 10\,GHz) 
For sources with typical spectral indices $\alpha \approx -0.7$, the point-source
sensitivity at the center of the 10\,GHz image is $\approx 2 \times$
more sensitive than that of the \citet{gm10} 1.4\,GHz image, which has $\theta_{1/2} = 1\,\farcs7$ resolution and rms noise $\sigma_\mathrm{n} \sim 4\,\mu\mathrm{Jy\,beam}^{-1}$.

We additionally made $(u,v)$-tapered images having 1\arcsec and
2\arcsec~synthesized beams to increase the brightness-temperature
sensitivity of our observations and investigate if our high-resolution
image missed significant numbers of extended sources.  The tapered
1\arcsec and 2\arcsec~images have an rms noises $\sigma_\mathrm{n}
\approx 1.1\,\mu\mathrm{Jy\,beam}^{-1}$ and $\sigma_\mathrm{n} \approx
1.5\,\mu\mathrm{Jy\,beam}^{-1}$, respectively, or roughly $2\times$
and $3\times$ higher than our full-resolution image.  However, the
corresponding brightness temperature rms values of the 1\arcsec and
2\arcsec~tapered images are $\sigma_\mathrm{n} \approx 13
\mathrm{\,mK}$ and $\sigma_\mathrm{n} \approx 4.7 \mathrm{\,mK}$, 
or  $\approx 10\times$ and $\approx 30 \times$ lower than
our full-resolution image, respectively.

\begin{center}
\begin{figure}
\epsscale{1.2}
\hspace*{-18pt}\plotone{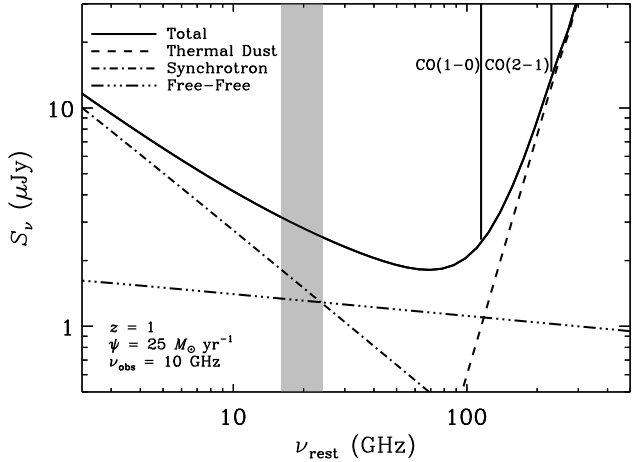}
\caption{A model radio spectrum of a $z = 1$ galaxy forming stars at a rate of $\psi \approx 25\,M_{\odot}\,$yr$^{-1}$.  
Synchrotron, free-free, and thermal dust emission components are identified, along with the $J=2\rightarrow1$ and $J=1\rightarrow0$ emission lines of CO.
The grayed region identifies the rest-frame bandpass of our 10\,GHz observations for such a source, illustrating that at $z\gtrsim1$ these data should become sensitive to free-free emission, and hence current star formation activity in high-redshift galaxies.  
}
\label{fig:sed}
\end{figure}
\end{center}

\subsubsection{Sensitivity of Deep Radio Imaging to Star Forming Galaxies at High-Redshift}
Coupling the relation between total infrared (IR; $8-1000\mu$m) luminosity and star formation rate in \citet[][Equation~15]{ejm12b}, which assumes a Kroupa \citep{pk01} initial mass function (IMF), with the locally measured IR-radio correlation \citep[i.e., $q_{\rm IR} = 2.64$;][]{efb03}, the star formation rate of a galaxy is related to its 1.4\,GHz spectral luminosity by 
\begin{equation}\label{eqn:radiosfr}
\psi(M_\odot \mathrm{\,yr}^{-1}) \approx 5.16 \times 10^{-29} 
\Biggl( \frac {L_\mathrm{1.4\,GHz}} {\mathrm{erg\,s^{-1}\,Hz}^{-1}} \Biggr)~.  
\end{equation}
At $z =1$, a $5\sigma_\mathrm{n} = 2.86\,\mu\mathrm{Jy\,beam}^{-1}$ 10\,GHz point source with a spectral index $\alpha = -0.7$ has a spectral luminosity $L_\mathrm{1.4\,GHz} \approx 4.85 \times 10^{29} \mathrm{\,erg\,s^{-1}\,Hz}^{-1}$, so Equation~\ref{eqn:radiosfr} gives a star-formation rate $\psi \approx 25\,M_{\odot}\,\mathrm{yr}^{-1}$.  

%\\
%assuming a Kroupa \citep{pk01} initial mass function (IMF).  \\
%{\bf OR}\\
% ejm -- decided to use our own recipe, since thought it would look funny to use someone else's... as if we don't have confidence in our previous work.  
%According to Equation~17 in \citet{ejm11b}, the star formation rate of a galaxy is related to its 1.4\,GHz spectral luminosity via the far-infrared--radio correlation by
%\begin{equation}\label{eqn:radiosfr}
%\psi(M_\odot \mathrm{\,yr}^{-1}) \approx 6.35 \times 10^{-29} 
%\Biggl( \frac {L_\mathrm{1.4\,GHz}} {\mathrm{erg\,s^{-1}\,Hz}^{-1}} \Biggr), 
%\end{equation}
%assuming a Kroupa \citep{pk01} initial mass function (IMF).  
%
%Taking the differing IMFs into account, this calibration is consistent with that presented in \citet{efb03}.  
%This assumes a non-thermal radio spectral index of $\alpha_{\rm NT}=-0.85$ (see \S\ref{sec:spx}).  
%At $z \approx 1$, a $5\sigma_\mathrm{n} = 2.86\,\mu\mathrm{Jy\,beam}^{-1}$ at 10\,GHz point source with a spectral index $\alpha = -0.7$ has a spectral luminosity $L_\mathrm{1.4\,GHz} \approx 4.85 \times 10^{29} \mathrm{\,erg\,s^{-1}\,Hz}^{-1}$, so Equation~\ref{eqn:radiosfr} gives a star-formation rate $\psi \approx 30\,M_{\odot}\,\mathrm{yr}^{-1}$.  

A model radio-to-far--infrared spectrum of such a galaxy is illustrated in Figure \ref{fig:sed}, indicating how these 10\,GHz observations are highly sensitive to the amount of free-free emission at $z\gtrsim1$, and hence current star formation.  
Furthermore, at the low radio flux densities achieved by our deep 10\,GHz imaging, the fraction of active galactic nuclei (AGN) detected relative to star-forming galaxies is extremely low \citep[e.g.,][]{jc84,s-cubed08}.  
In Figure~4 of \citet{s-cubed08}, the AGN fraction at several frequencies are shown including 1.4, 4.8, and 18\,GHz, where the lowest AGN fraction is indeed found at the highest frequency of 18\,GHz, being $\lesssim10$\%.    %the highest frequencies.  
Consequently, by being sensitive to free-free emission from galaxies forming stars at a rate of $\psi \gtrsim 25\,M_{\odot}\,\mathrm{yr}^{-1}$ at $z\sim1$, our 10\,GHz data are sensitive to galaxies that contribute to roughly half of the cosmic star formation rate density at these epochs \citep[e.g.,][]{ejm11a,bm13}.
%%REF1
Furthermore, given this star formation rate threshold, our data reach a mass-limit where the $z\sim 1$ cosmic star formation rate density peaks, and hence down to the most representative star forming sources at this epoch \citep{ak11}.  %Karim et al. 2011).

\subsection{Source Finding and Photometry \label{sec:srcfind}}
We used the {\sc PyBDSM}\footnote{\url{http://www.astron.nl/citt/pybdsm/}}
\citep{pybdsm} source detection package to locate and measure the
integrated flux densities of the radio sources in each of our images.
To do this, {\sc PyBDSM} identifies ``islands" of contiguous pixels
above a detection threshold and fits each island with Gaussians.  We
ran {\sc PyBDSM} over the entire image using the standard default
setting, except that we lowered the pixel threshold for keeping fitted
sources from 5 to $3.5 \sigma_\mathrm{n}$.  We adopted a minimum
threshold of $3.5 \sigma_\mathrm{n}$ as we consider such sources
having optical identifications potentially significant.  A primary
beam correction was then applied to the reported peak brightnesses and integrated
flux densities (along with their errors) using the frequency-dependent primary beam
correction at 10\,GHz given in EVLA Memo\#\,195 (Perley
2016)\footnote{\url{https://library.nrao.edu/public/memos/evla/EVLAM-195.pdf}}.
For this paper, we kept sources out to a radius of $\approx
3\,\farcm9$ from the phase center, where the sky response
drops to $\approx5$\% of the on-axis value.  
% and a reliable flux density can still be recovered.  
%%REF1
For reference, the HWHM of the primary beam at 10\,GHz is $\approx 2 \farcm 125$.  
%%REF1
%XX JJC: The numbers below seem backwards.  Should there be 114 above 5 sigma
% and 1412 above 3.5 sigma?
A total of 1412 and 114 sources were ``detected" (see \S\ref{sec:fd}) above a threshold of 3.5 and $5 \sigma_\mathrm{n}$, respectively.  
To assess the reliability of detections as faint as $3.5 \sigma_\mathrm{n}$, we used existing deep {\it HST} and {\it Spitzer} data to identify OIR counterparts.
%XX JJC: Eric: The referee asked us to justify 3.5 sigma detections, and you noted
%XX JIM: PLEASE COMMENT on your respones.  I must admit I'm still worried about 
%XX the reliability of the 13 sources 
%XX with 3.5 sigma < S < 5 sigma and OIR IDs within 0.3 arcsec.
%XX Inserting rho = .072 arcsec^{-2} into Equation 6 says that
%XX there should be about 26 false IDs among 1298 sources.  
%XX If we believe in our 13 IDs, it's only because they are relatively bright.
%XX We should say so, choose a magnitude cutoff, and recalculate the reliability
%XX of the sources with bright IDs.

In Table \ref{tbl-1} we list the deconvolved source parameters from
{\sc PyBDSM} for detections from the full-resolution image that are
confirmed by having OIR counterparts (see \S\ref{sec:optID}).  
Sources are split by their 10\,GHz detection confidence, being either at a
significance of $ S_\mathrm{P}/\sigma_\mathrm{n} \geq 5$  or $3.5 \leq S_{\rm
  P}/\sigma_\mathrm{n} < 5$.
%XX jjc I replaced $\sigma_\mathrm{P}$ by $\sigma_\mathrm{n}$
%XX jjc in Table 1 to be consistent with the text and because it is
%XX jjc the map noise level, not the formal uncertainty in the fit. 
%XX jjc I also reformatted some table columns for better vertical alignment. 
It is worth emphasizing that images from the full survey, which will include
multiple pointings per source, will be able to confirm any
questionable sources without OIR counterparts detected in these
single-pointing pilot observations.  Listed parameters include the
source positions, peak brightnesses ($S_{\rm P}$),
integrated flux densities ($S_{\rm I}$), 
best estimates for the total flux densities ($S_{\rm *}$) and deconvolved FWHM source
sizes ($\theta_{\rm M} \times \theta_{\rm m}$).
Deconvolved source sizes were calculated such that 
\begin{equation}
\label{eq:deconv}
\theta = \sqrt{\phi^2 - \theta_{1/2}^2}~, 
\end{equation}
where $\phi$ is the FWHM of the fitted major or minor axis.  
The uncertainties in the fitted parameters include the effects of correlated noise in synthesis images \citep{jc97}.  
Uncertainties on the deconvovled source sizes were calculated as follows 
\begin{equation}
\label{eq:deconvunc}
\left(\frac{\sigma_{\theta}}{\sigma_{\phi}}\right) = \left[1 - \left(\frac{\theta_{1/2}}{\phi}\right)^{2}\right]^{-1/2}~.  
\end{equation}
%XX jjc The text states that the radio position errors are << 0.1 arcsec,
%XX jjc but Table 1 lists them with only 0.1 arcsec precision.  We should
%XX jjc add at least one more digit of precision to both RA and Decl.

For a few sources, {\sc PyBDSM} fit multiple Gaussian components,
typically including a very narrow component at the peak of the source.
One instance of this is for the known FR\,I radio galaxy at $z=1.013$ \citep{richards98}, where we are able to resolve some of its jet structure.  
Because we are interested in measuring the extent of star-forming galaxy disks, we used {\sc imfit} in CASA to fit single Gaussians for these cases and report the corresponding deconvolved source parameters.

%XX jjc The text is ambiguous about how the upper limits to the deconvolved
%XX jjc source sizes were assigned.  Is the upper limit
%XX jjc UL = (beam + 2 sigma) - beam   
%XX jjc or
%XX jjc UL = sqrt( (beam + 2 sigma)^2 - beam^2 )
%XX jjc The latter is the correct way.
% ejm: UL's were caltulated in the correct way.  Included an equation to remove ambiguity.  
%
%For sources whose fitted major or minor axes are less than 2$\sigma$
%larger than the synthesized beam, we assigned corresponding 2$\sigma$
%upper limits to the deconvolved source sizes such that 
%\begin{equation}
%\theta_{\rm UL} = \sqrt{(\phi + 2\sigma_{\phi})^{2} - \theta_{1/2}^{2}}.  
%\label{eq:sizeul}  
%\end{equation}
%And, for those sources whose major axes are unresolved, 
%we take their total flux density to be the geometric mean of the peak and integrated
%flux densities reported by {\sc PyBDSM} \citep{jc97}. 
%XX jjc Cite Condon, J. J. 1997, PASP, 109, 166 to justify using the
%XX jjc geometric mean.
Individual sources are considered to be confidently resolved if \(\phi_{\rm M} - \theta_{1/2} \geq 2\sigma_{\phi_{\rm M}}\), and are identified in Tables~\ref{tbl-1} and \ref{tbl-2}.  
For instances where {\sc PyBDSM} reported unphysical fitted sizes, being equal to or smaller than the synthesized beam, a size of 0 in listed in Tables~\ref{tbl-1} and \ref{tbl-2} where the associated uncertainty corresponds to the 1$\sigma$ upper limit of the deconvolved source size.  
For sources whose major axes are resolved, the integrated flux densities from the source fitting are taken as the best estimate for the sources total flux density ($S_{*}$).  
For sources whose fitted major axes are less than 2$\sigma_{\phi_{\rm M}}$ larger than the synthesized beam, 
we instead take their total flux density to be the geometric mean of the peak brightness and integrated flux densities reported by {\sc PyBDSM} \citep{jc97}.

%XX jjc I also tweaked Table 2.
In Table \ref{tbl-2}, we similarly list the deconvolved source
parameters and photometry from the 1\arcsec~$(u,v)$-tapered image, or
from the 2\arcsec~image in the three cases where the total flux density
in the 2\arcsec~image is $>3 \sigma$ larger than in the
1\arcsec~image.  
The flux densities recovered in the 2\arcsec~tapered image for these three sources are larger by factors of $\approx 1.1$, 2.9, and 1.2 in order of appearance in Table~\ref{tbl-2}.  %$1.1-2.6$ with a median of 1.6.

\input{tab1}
%\input{tab1.jc}

%\hspace*{-20pt}
%\input{tab2}
%\input{tab2.jc}

\subsubsection{Optical Identification \label{sec:optID}}
We identified our radio sources with their OIR counterparts in
\citet{im16} by position coincidence using the criteria presented in
the Appendix of \citet{con75}.
%XX Eric: add to your bibliography:
%XX \bibitem[Condon et al.(1975)]{con75} Condon, J.~J., Balonek, T.~J.,
%XX \& Jauncey, D.~L. 1975, AJ, 80, 887 In the VLA astrometric frame
The radio position accuracy is noise limited, so the one-dimensional
radio position errors parallel to the fitted FHWM major axis
$\phi_\mathrm{M}$ and minor axis $\phi_\mathrm{m}$ are Gaussian
distributed with rms
\begin{equation}\label{eqn:sig1d}
\sigma_\mathrm{M} \approx  \frac{\phi_\mathrm{M}}{\sqrt{8 \ln 2}} 
\biggl(\frac{\sigma_\mathrm{P}}{S_\mathrm{P}}\biggr)  \mathrm{\quad and \quad}
\sigma_\mathrm{m} \approx  \frac{\phi_\mathrm{m}}{\sqrt{8 \ln 2}} 
\biggl(\frac{\sigma_\mathrm{P}}{S_\mathrm{P}}\biggr) ,
\end{equation}
respectively, where ($S_\mathrm{P}/ \sigma_\mathrm{P} \approx
S_\mathrm{P} / \sigma_\mathrm{n}$) is the fitted
peak signal-to-noise ratio ($SNR$).  Even $SNR = 5$ point sources have
very small $\sigma_\mathrm{M} \approx \sigma_\mathrm{m} \approx
0\,\farcs 019$.  Nearly all of the OIR identifications have very high
$SNR$, so their rms uncertainties $\sigma_\mathrm{OIR}$ are dominated
by systematic differences between the OIR and radio reference frames.
We made preliminary identifications and found systematic OIR minus
radio offsets $\Delta\alpha = +3\,$mas and $\Delta\delta = +30\,$mas.
After these offsets were removed, the remaining OIR errors have zero
mean and rms $\sigma_\mathrm{OIR} \approx 0\,\farcs 025$.

If the radio and OIR sources coincide exactly on the sky, their measured
radio-optical offsets $r$ should have a Rayleigh distribution
\begin{equation}
P(r) = \frac {r} {\sigma^2} \exp \biggl( - \frac{r^2}{2 \sigma^2} \biggr)
\end{equation}
with 
\begin{equation}\label{eqn:rmsoffset}
\sigma \approx \sqrt{ \left( \frac{\phi_{\rm M}}{\sqrt{8 \ln{2}}}
\frac{\sigma_\mathrm{n}}{S_{\mathrm P}} \right)^{2} + \sigma_{\rm OIR}^{2} }~.
\end{equation}

\begin{figure}[t]
\epsscale{1.2}
\plotone{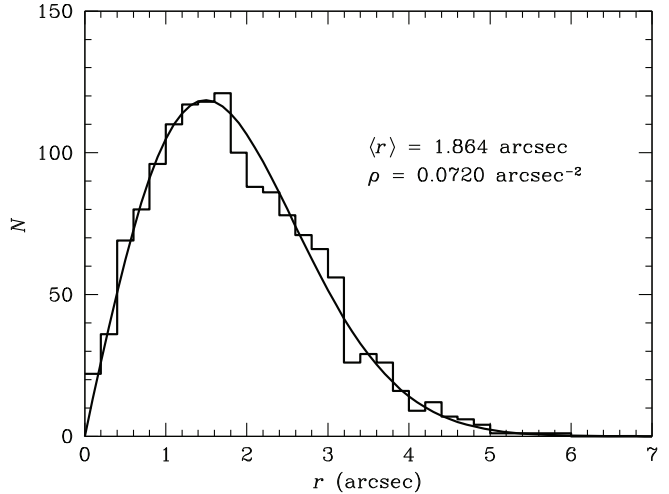}
\caption{This histogram showing the offsets $r$ of OIR candidates from
  1454 random positions in the radio image is well matched by
  the Rayleigh distribution expected for an OIR sky density $\rho =
  0.0720 \mathrm{~arcsec}^{-2}$.
\label{fig:neighbor}}
\end{figure}

The probability distribution of the angular distance $r$ 
%XX JJC: I reversed the next line for clarity:
from a radio source to the nearest unrelated optical object  
is
\begin{equation}\label{eqn:falseid}
P(r) = 2 \pi \rho r \exp(- \pi \rho r^2) ~,
\end{equation}
where $\rho$ is the sky density of optical identification candidates.
The mean angular distance to the nearest unrelated optical object is
\begin{equation}
\langle r \rangle = \frac {1} {2 \sqrt\rho}~.
\end{equation}
Figure~\ref{fig:neighbor} shows the distribution of angular
distances $r$ to the OIR sources nearest to a large sample of random
positions in our fields (histogram) and the best-fit Rayleigh
distribution, which implies $\rho \approx 0.0720
\mathrm{~arcsec}^{-2}$.  

To avoid incorrect optical identifications, it is necessary to keep
$\rho \sigma^2 \ll 1$. 
The quantity $k \equiv (1 + 2 \pi \rho \sigma^2)$ measures the identification
candidate sky density in units of $\sigma^2$, and $m \equiv r_\mathrm{s} /
\sigma$ describes the search radius $r_\mathrm{s}$ in units of the rms
position error.  For $\rho \approx  0.0720 \mathrm{~arcsec}^{-2}$ and
$\sigma \approx  0\,\farcs 03$, $k \approx 1.00041$. 

The final quantity needed to calculate the completeness $C$ and
reliability $R$ of position-coincidence identifications is the
fraction $f$ of sources that do have OIR counterparts brighter than
some magnitude cutoff.  The resulting identifications should have
completeness
\begin{equation}
C = \frac{1 - \exp(- k m^2 / 2)}{k}
\end{equation}
and reliability
\begin{equation}
R = C \biggl[ \frac{1}{f} + \biggl( 1 - \frac{1}{f} \biggr)
\exp [ m^2(1-k)/2] - \exp(-m^2 k / 2)\biggr]^{-1}~.
\end{equation}
The best choice of the free parameter $m$ limiting the search radius
is a compromise between high completeness (large $m$ ensures that true
associations are not overlooked) and high reliability (low $m$ avoids
incorrect identifications with unrelated OIR objects nearby on the
sky); it usually lies in the range $2 < m < 3$.  The value of $k$ in
our sample is so small that we can safely choose $m = 3$.  This
ensures high completeness $C \approx 0.989$, and high reliability $R >
0.983$ for any value of $f > 0.1$.

\begin{figure}[htb!]
\epsscale{1.2}
\plotone{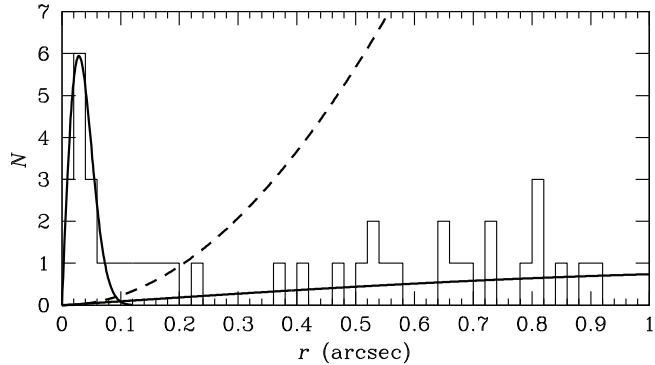}
\caption{
The histogram shows the numbers $N$ of radio sources per bin of width
$\Delta r = 0\,\farcs02$ with nearest OIR neighbors at angular offsets $r
< 1''$.  The sharp Rayleigh distribution matches the observed 14
sources identified by position coincidence with $m < 3$ and median
$\langle \sigma \rangle  \approx 0\,\farcs03$.  The broad curve is
the expected differential distribution for unrelated OIR objects with sky
density $\rho = 0.0720 \mathrm{~arcsec}^{-2}$, 
and the dashed curve is the calculated
cumulative distribution $N(>r)$.  It predicts one unrelated optical
object will fall
within $r = 0\,\farcs 2$ of a $5\sigma_{\rm n}$ radio source.
\label{fig:idsep}}
\end{figure}

The histogram in Figure~\ref{fig:idsep} shows the distribution of
radio/OIR offsets $r < 1''$ for all 114 radio sources with $SNR
\geq 5$ in our high-resolution ($0\,\farcs22$) image.  The sharp
Rayleigh distribution cutting off near $r = 0\,\farcs 09$ fits the $N
= 14$ identifications with $m < 3$ and $\langle \sigma \rangle \approx
0\,\farcs03$. The slowly rising solid curve indicates the number of
unrelated OIR objects with sky density $\rho = 0.0720 (114 - 14) / 114
\mathrm{~arcsec}^{-2}$ expected per bin of width $\Delta r = 0\,\farcs
02$, and the dashed curve is their calculated cumulative distribution
$N(>r)$.  The observed density of unrelated OIR objects matches the
calculated curves quite well for $0\,\farcs 2 < r < 1''$, 
indicating that clustering on scales $\lesssim1\arcsec$ (i.e., $\lesssim 8\,$kpc at $z\gtrsim1$) does not detectably increase the number of nearby OIR objects.  
However, there are five ``unexpected'' OIR objects with $0\,\farcs 09 \lesssim r
\lesssim 0\,\farcs2$ where we expected only one in our sample of 114
radio sources.  This discrepancy reveals that the OIR
identifications of some faint radio sources may genuinely be slightly
offset in position; in particular, the most intense radio emission may
come from a dusty star-forming region in a merging system with patchy
OIR obscuration.  For example, an $r \approx 0\,\farcs6$~(4\,kpc)
separation was measured between rest-frame UV and far-infrared
emission peaks in the high-redshift starburst galaxy GN20
\citep{jh15}.  We inspected the OIR images of the five ``unexpected''
objects and estimate that four of the five sources in Table~\ref{tbl-1} with
$S_\mathrm{P} / \sigma_\mathrm{n} \geq 5$ and $ 0\,\farcs 09 < r <
0\,\farcs 2$ are correctly identified.
 
We carried out a similar analysis to asses the reliability of sources between $3.5 \leq S_\mathrm{P} / \sigma_\mathrm{n}  < 5$, for which there was a total of 544 sources in the relative gain-weighted beam solid angle.  
 For a $SNR = 3.5$, Equation~\ref{eqn:rmsoffset} gives an rms position error of $\sigma=0\,\farcs37$.  
 Setting $m=3$, we estimate a total of 1.5 false detections within a search radius of $r < 0\,\farcs11$.  
 In Table~\ref{tbl-1} there are 3, $3.5 \leq S_\mathrm{P} / \sigma_\mathrm{n}  < 5$ sources with $0\,\farcs11 < r < 0\,\farcs2$ that we consider to be reliable since: 
 one is associated with a bright $JH_{\rm NIR} = 20.61$\,mag galaxy, 
 another is additionally detected at 1.4\,GHz, 
 and the third is is associated with a heavily obscured, morphologically-disturbed galaxy for which a large offset appears physical based on a visual inspection.

%XX jjc The three leftmost panels are labeled 10Hz instead of 10 GHz.
\begin{center}
\begin{figure*}
%\epsscale{1.35}
\hspace*{-18pt}\plotone{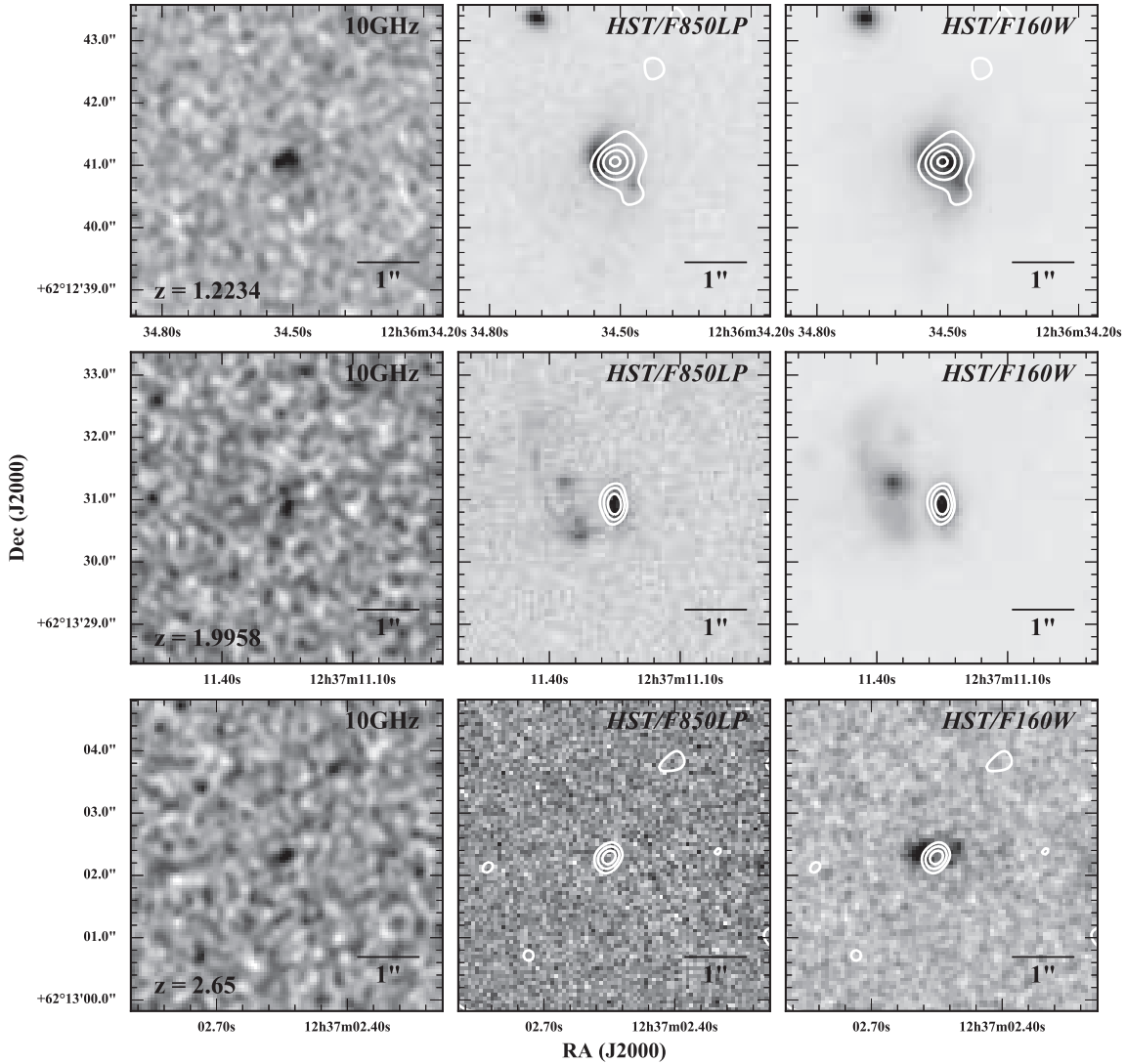}
\caption{Examples of sources detected in our 10\,GHz pilot
  observations.  Also shown are {\it HST}/ACS $z$-band and {\it HST}/WFC3 F160W images with 10\,GHz
  contours overlaid.  Each panel is a $5\arcsec\times5\arcsec$~cutout.
  Along with redshifts, 1\arcsec~scale bars are shown.  The top row
  illustrates our ability to resolve the two cores in this merging
  galaxy pair at $z=1.2234$, while the bottom row illustrates our
  ability to detect optically invisible sources in the radio at high redshifts.  }
\label{fig:srcs}
\end{figure*}
\end{center}

We report a total of 32 reliable identifications for radio sources
detected at $\geq3.5\sigma_\mathrm{n}$ significance in our
full-resolution 10\,GHz image.  The median positional uncertainty among
these 32 sources is $\langle 3 \sigma \rangle \approx0\,\farcs1$, which is typically larger than the median measured separation between the OIR and radio positions $\langle r \rangle \approx 64\pm11$\,mas.
However, there are 9 instances ($\approx 30$\%) where the radio and OIR separation exceeds 3$\sigma$ by a median of $\approx 53\pm12\,\mathrm{mas} \lesssim 0.41\pm0.10\,\mathrm{kpc}$ at any redshift.  The median distance
between the radio and OIR centroids for the 9 offset identifications is $149\pm13\,\mathrm{mas} \lesssim 1.17\pm0.10\,\mathrm{kpc}$.

In the 1\arcsec~resolution image we report a total of 27 reliable
identifications detected at $\geq3.5\sigma$ and matched to an OIR
counterpart, 6 of which are not detected in the full-resolution image.
The median positional uncertainty among these 27 sources is $\langle 3 \sigma \rangle \approx0\,\farcs3$, which is typically larger than the measured separation between the OIR and radio positions, having a median separation of $\approx 126\pm19$\,mas.  
In this image, the radio and OIR separation for 4 sources is found to be larger than the 3$\sigma$ positional uncertainty by a median value of $\approx31\pm13$\,mas, or $\approx 262\pm102$\,pc at their distances.  
In other terms, the median distance between the radio and OIR centroids for these 4 sources is $188\pm66$\,mas, or $\lesssim 1.6\pm0.4$\,kpc at any redshift.

In Tables~\ref{tbl-1} and \ref{tbl-2}, spectroscopic redshifts (\citealp{jcohen00,gw04, ams04, tt05, nr06, bcw08,dtf08, ht11}; D. Stern et al.\ in preparation; this paper) are given along with grism and photometric redshifts included in \citet{im16}.  
Additional notes on specific redshifts can be found in Appendix \ref{sec:AppendixA}.  
%For one source having coordinates $12 37 02.539$~$+$+62 13 02.32$ the photometric redshift reported in \citet{im16} is $z = 5.0385$, however we report the photometric redshift included in D. Kodra et al. (in preparation, $z=2.650$) as we believe it to be more reliable.    
In every case, we provide the angular separation $r$ between the 10\,GHz detection and the OIR counterpart plus the NIR magnitude ($JH_{\rm NIR}$) that is derived from some combination of $J_{125}$, $JH_{140}$ and $H_{160}$ {\it HST}/WFC3 images scaled to the $JH_{140}$ AB zeropoint as described in \citet{im16}.   We present image cutouts for 3 examples of 10\,GHz sources matched to {\it HST} counterparts in Figure \ref{fig:srcs}, one of which illustrates how dust obscuration can cause a statistically significant offset between the radio and OIR positions.

%is measured to be $\approx0\,\farcs0$ in the X-band and $\approx$0\,\farcs025 in the {\it HST} data \citep{im16}.  
%However, when matching to existing OIR data we additionally include a padding of 0\,\farcs6 to the final search radius used per source, as the physical separation of OIR and highly obscured radio/FIR emission has been found to be this large in high redshift starbursts such as GN20 \citep{jh15}).  
%This additional factor dominates the search radius.  
%{\bf From Gabe:  }Given that the optical source density is $\approx$220\,arcmin$^{-2}$ (1 source per every $]approx$16 arcsec$^2$), this additional padding in search radius is not expected to lead to a significant number of false detections.  
%{\bf WE MEAURE:  }{Given that the optical source density is $\approx$260\,arcmin$^{-2}$ (1 source per every $]\approx$14 arcsec$^2$), this additional padding in search radius is not expected to lead to a significant number of false detections.  

\subsubsection{A Warning About 5$\sigma_\mathrm{n}$ ``Detections" in Sparse Images \label{sec:fd}}

\begin{figure}
\epsscale{1.2}
\plotone{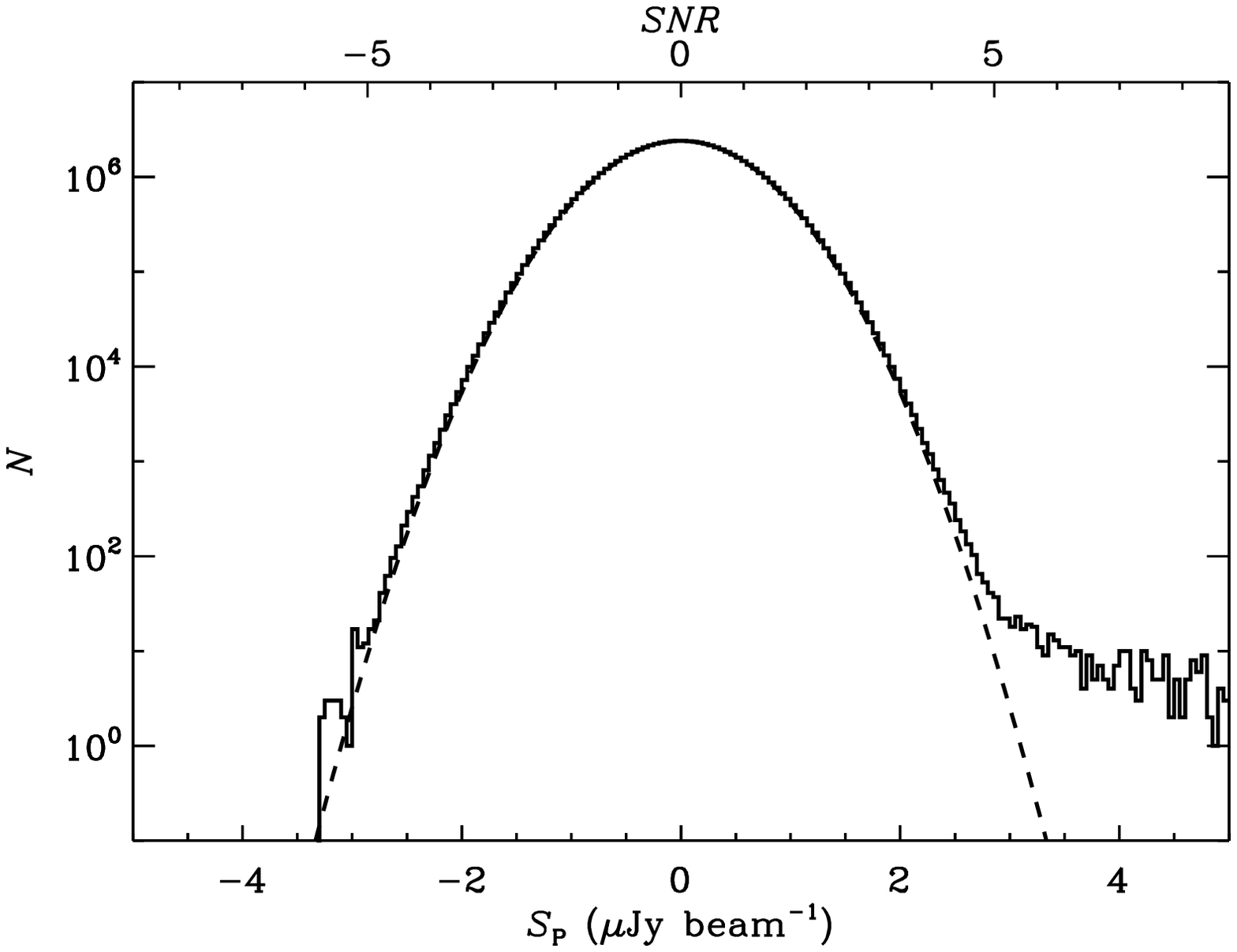}
\caption{The pixel brightness distribution of the high-resolution
  image uncorrected for primary-beam attenuation, over the entire
  10\,GHz primary beam area used for our source identifications.
The bin width is  a
  $0.05\,\mu \mathrm{Jy\,beam}^{-1}$.  The dashed line is a
  Gaussian fit with rms $\sigma = 0.572\,\mu\mathrm{Jy\,beam}^{-1}$
  (i.e., the rms measured in the image out to 5\% of the primary beam
  response).  The noise in the 10\,GHz image appears Gaussian and
  extremely well behaved.  }
\label{fig:noise}
\end{figure}

There are 114 sources with $SNR \geq 5$ in our image covering $\Omega = 3.6 \times 10^5
\mathrm{~arcsec}^2$ with $\theta_{1/2} = 0\,\farcs22$ FWHM resolution.  The
noise has the same angular power spectrum as the synthesized beam, so
there are $N_\mathrm{n} \approx 8 \ln(2) \Omega / (\pi \theta_{1/2}^2)
\approx 1.3 \times 10^7$ independent noise samples in our image.  The
image noise amplitude distribution (Figure~\ref{fig:noise}) is nearly, 
but not perfectly, Gaussian.  If it were perfectly Gaussian, the
probability that any sample would exceed $+5 \sigma_\mathrm{n}$ is
only $P(> +5\sigma_\mathrm{n}) = 2.87 \times 10^{-7}$ and there would
be only $N_\mathrm{n} P(> +5\sigma_\mathrm{n}) \sim 4$ false radio
sources stronger than $+5 \sigma_\mathrm{n}$.  
However, we believe that nearly all of the 95 optically unidentified $5\sigma_\mathrm{n}$
sources are actually spurious for several reasons:
\noindent(1) The histogram of the $SNR$s for the 95 unidentified sources is 
extremely steep and cuts off below $SNR =  6.2$.  
\noindent (2) We cannot match any of these 95 sources reliably to sources in the extremely deep {\it Spitzer}/IRAC data at 3.6 or
4.5\,$\mu$m within a 1\arcsec~radius using catalogs compiled from
\citet{de11}, which is surprising if these sources were in fact real
but within heavily obscured galaxies.
\noindent (3) The 95 unidentified sources are uniformly distributed across the primary beam as one would expect for random noise; they are not concentrated toward the center of the primary beam where 
%JXX JC: I replaced 'the noise is minimized as one would expect 
%XX for real sources.' by
real sources are stronger.
 \noindent (4) We found a similar number (92) of sources more
negative than $-5\sigma_\mathrm{n}$ in our high-resolution image by
multiplying the image intensity units by $-1$ and running PyBDSM to
find negative sources.

%XX JJC: Why are will still quoting 554 nJy/beam (e.g., in the abstract)
%XX when the rms noise inside the  primary beam is 572 nJy/beam?
%XX The figure caption below uses the correct 572 nJy/beam.
%jjc It appears that the rms noise in the high-resolution image
%jjc is 572 nJy/beam inside the 5% primary beam circle, yet most
%jjc of the text quotes 554 nJy/beam.  If the 554 nJy/beam comes
%jjc from including the edges of the image that we don't use
%jjc then maybe we should be using 572 nJy/beam throughout the paper.

We believe that the false $\pm 5\sigma$ ``detections" in our
high-resolution image are simply the result of faint non-Gaussian
image fluctuations related to unedited RFI, imperfectly cleaned
sidelobes, and calibration errors.  The lesson here is that sources in
very sparse images (many clean beam solid angles per source) cannot be
trusted at the $5\sigma_\mathrm{n}$ level, even if the image noise
looks perfectly uniform and Gaussian to the human eye.

Low-resolution images covering fewer beam solid angles or
low-frequency images containing larger numbers of sources should yield
more reliable $5\sigma_\mathrm{n}$ sources.  The situation is
drastically different for the case of the 1\arcsec resolution image, in
which we detected a total of 15 sources with $S_\mathrm{P}/\sigma_\mathrm{n} \geq 5
$ confidence and all but one is reliably matched to
an optical counterpart.  The single source that is not reliably
matched is $r \approx 0\,\farcs65$ from its nearest OIR counterpart.
However, this same OIR counterpart (located at a redshift of $z =
2.004$) is reliably matched to a radio counterpart in the $0\,\farcs
22$ resolution image and is offset by only $r \approx 0\,\farcs 093$.

%\subsection{Spectral Indices and Estimated Thermal Fractions \label{sec:spx}}
\subsection{Using Spectral Indices to Estimate Thermal Fractions \label{sec:spx}}

%XX JJC: Both Sections 2.4 and 3.3 are titled 'spectral indices and thermal fractions',
%XX  which may confuse the reader and may have confused the referee, who thought
%XX  they were redundant.  Section 2.4 covers only HOW we estimated thermal
%XX  fractions from spectral indices, while Section 3.3 actually discusses the results.
%XX Maybe the title for Section 2.4 should be changed to something like
%XX  ``Using spectral indices to estimate thermal fractions''

Table~\ref{tbl-2} lists the radio spectral indices $\alpha$ between
1.4 and 10\,GHz of sources found by matching our 10\,GHz positions
with sources in the 1.4\,GHz \citet{gm10} catalog.  
Their 1.4\,GHz images have $1\farcs7$ resolution and rms noise $\sigma_\mathrm{n}
\approx 4\,\mu\mathrm{Jy\,beam}^{-1}$.  
Of the 27 sources detected in the 1\arcsec-resolution image, we were able to find 1.4\,GHz counterparts for 19 of them ($\approx 70$\%).  
For the 10\,GHz sources not having 1.4\,GHz counterparts, Table~\ref{tbl-2} lists lower limits to
$\alpha$ calculated using a 1.4\,GHz upper limit of $S_\mathrm{P} < 5 \sigma_\mathrm{n} \approx
19.5\,\mu\mathrm{Jy\,beam}^{-1}$.  If we assume all sources have the
same non-thermal spectral index, we can use the measured spectral
indices to estimate the fractional contributions from thermal emission
at the rest-frame frequencies $(\nu_{\mathrm{r}}/\mathrm{GHz}) =
10(1+z)$ for sources having known redshifts $z$ \citep[e.g.,
][]{kwb84,ejm12b}.  These are given in Table \ref{tbl-2}, placing
limits where necessary.  This simple thermal decomposition is
sensitive to the estimated non-thermal spectral index, assumes that
the free-free emission is optically thin at rest-frame frequencies
$(\nu_{\rm r}/{\rm GHz}) \gtrsim 1.4(1+z)$
\citep[e.g.,][]{tm10}, and that there is an insignificant contribution
of both anomalous microwave emission at $\sim$33\,GHz
\citep[e.g.,][]{ejm10} and thermal dust emission in the rest frequency
range $10 \lesssim \nu \lesssim 40$\,GHz.  We took the non-thermal
spectral index to be $\alpha_{\rm NT} = -0.85$, which is the
average non-thermal spectral index found among the 10 star-forming
regions studied in NGC\,6946 by \citet{ejm11b} and very similar to the
average value found by \citet[][$\langle \alpha_{\rm NT}
  \rangle = -0.83$ with an rms scatter of $\sigma = 0.13$]{nkw97}
globally for a sample of 74 nearby galaxies.  To those sources having
measured spectral indices $\alpha > \alpha_{\rm NT} = -0.85$ we 
assigned non-thermal indices $\alpha_{\mathrm{NT}} = \alpha - 0.1$.

%XX jjc The median redshifts are all quoted with 3 digits precision even
%XX jjc though their rms errors are > 0.1.  Maybe we should use only 2
%XX jjc digits; e.g., 1.24 \pm 0.12 instead of 1.241 \pm 0.116.
% ejm:  Agreed.  Done. 

\begin{figure}
\epsscale{1.2}
\plotone{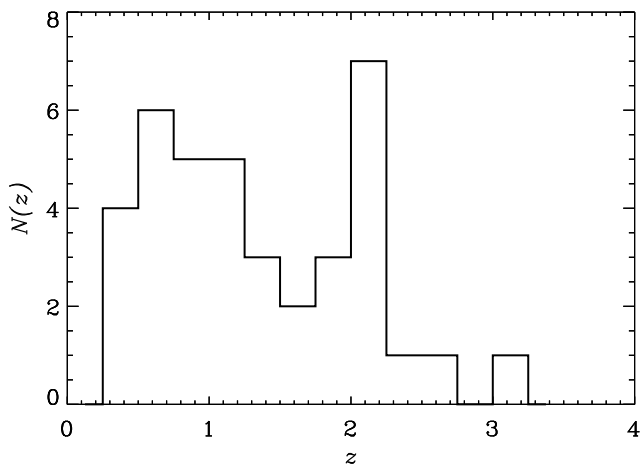}
\caption{The redshift distribution of all 38 unique sources detected
  in the full-resolution and/or 1\arcsec~tapered 10\,GHz images with
  $S_\mathrm{P} \geq3.5 \sigma_\mathrm{n}$ and confirmed by having OIR
  counterparts.  Their median redshift is $\langle z \rangle =
  1.24\pm0.15$.  
  An excess of sources (7 out of 38) are found in the narrow range of $1.9958 \leq z \leq 2.0180$, and are likely members of an over-density traced by SMGs, optically-faint radio sources, and other star-forming galaxies \citep{sc09}.  
}
\label{fig:zdist}
\end{figure}

\section{Results and Discussion}

In this section we present the results 
%%REF1
for a flux-limited sample of galaxies 
from this pilot X-band imaging program of GOODS-N, taking advantage of both the high angular resolution delivered by the VLA A-configuration, 
%XX JJC: I hope we filled the A-configuration hole with C-configuration data. Otherwise, the
%XX  ``bowl'' caused by missing short spacings will cause us to underestimate the angular
%XX sizes and total flux densities of extended sources, even in the tapered images.
along with its centrally concentrated core of dishes, allowing us to make images with various $(u,v)$-weightings to improve the brightness temperature sensitivity of the data.
As stated in \ref{sec:imging}, the inclusion of the C-configuration data with our imaging helps to fill in the hole in the $(u,v)$-plane left by the A-configuration data alone that, if not accounted for, will result in underestimated angular sizes and integrated flux densities of extended sources even in the 1\arcsec~and 2\arcsec~tapered images.  

\subsection{Redshift Distributions}

The $0\,\farcs22$ full-resolution 10\,GHz image contains 32 reliable sources at $\geq3.5\sigma_{\rm n}$ having OIR counterparts with measured redshifts.
Their median redshift is $\langle z \rangle = 1.24\pm 0.16$.  
%The two apparently most distant sources in the 10\,GHz-detected sample have photometric redshifts only, without spectroscopic confirmation (i.e., $z_{\rm phot} = 3.24$ and 5.04).  
%Even if these are later found to be at much lower redshift, the statistics of our redshift distribution will not change significantly, nor any of the results and conclusions presented in this paper.    
The 1\arcsec-resolution image contains 27 sources with OIR counterparts
and measured redshifts.  The median redshift of these sources is
slightly lower: $ \langle z \rangle = 1.01\pm0.16$, most likely
because the 1\arcsec~tapered image is less sensitive to point sources
but more sensitive to extended sources, which will tend to be at lower
redshifts.  
% EJM: Not a bad idea.  And looking at teh z-distribution for just resolved source you can see the peak shift to lower z, but that is sort of just re-stating the median shifting.  
%XX JJC: Should we distinguish the extended sources (by cross-hatching, for example) from
%XX  the point sources in this redshift histogram, to justify the preceding sentence?
Figure \ref{fig:zdist} shows the redshift distribution for
all 38 unique sources detected in the full-resolution and/or
1\arcsec~tapered images.  The median redshift of all sources with OIR
counterparts is $\langle z \rangle = 1.24\pm0.15$.
Figure \ref{fig:zdist} also shows an excess of sources (7 out of 38) having redshifts in the narrow range spanning $1.9958 \leq z \leq 2.0180$.  
These sources are likely members of an over-density traced by sub-mm galaxies (SMGs), optically-faint radio sources, and other star-forming galaxies \citep{sc09}.  
%%REF1
In fact, at least 6 of the 38 unique 10\,GHz-detected sources reported here are 850\,$\mu$m-detected SMGs \citep{ap05,sc09}, 4 of which appear to be members of this $z\approx2$ over-density.

\begin{figure}
\epsscale{1.2}
\plotone{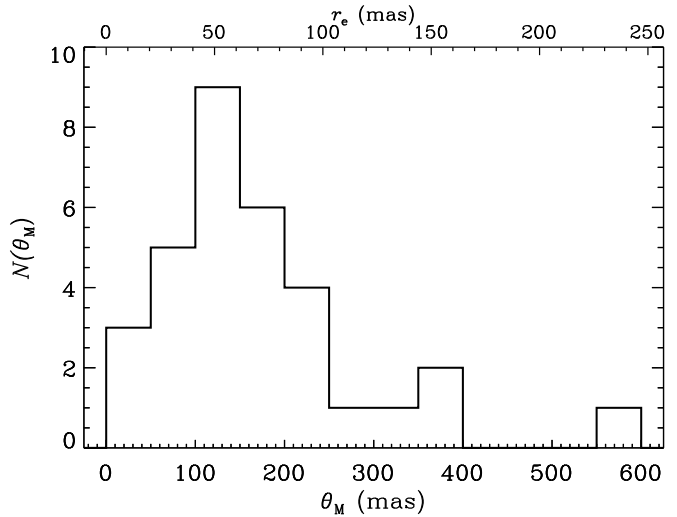}
\caption{The distribution of deconvolved source FHWM major axes
  $\theta_\mathrm{M}$ for all 32 sources detected in the
  full-resolution image.  The corresponding effective radii $r_{\mathrm
    e}$ are marked on the upper abscissa.  The weighted median angular size is
  $\langle \theta_{\mathrm M} \rangle = 167\pm32$\,mas
  and the weighted rms scatter in $\theta_\mathrm{M}$ is 91\,mas.  The
  corresponding median effective radius is $\langle r_{\mathrm e}
  \rangle = 69\pm 13$\,mas and the rms scatter in $r_\mathrm{e}$ is
  38\,mas.  }
\label{fig:sizedist}
\end{figure}

%XX JJC:  The referee suggested ``using present tense throughout the results section
%XX  (e.g., p. 9 ``plotted'', p. 10 ``compared'', and p. 12 ``detected''.''
%XX If something happened only in the past, it should be in past tense, newspaper
%XX headlines notwithstanding.  I searched for ``plotted'' in this .tex file,
%XX and all examples (e.g., in the caption below) should be in past tense.
%XX In general, we plotted something in the past, and we aren't plotting something now.
%XX Likewise for ``compared''; we made the comparisons in the past and are now just
%XX writing about them.  Our referee is clearly not a native English speaker, and
%XX English has a lot more tenses than most other languages (nearly everything in
%XX Chinese is in the present tense), so I think he/she is just wrong about using
%XX the present tense in these cases.
\begin{figure}
\epsscale{1.2}
\plotone{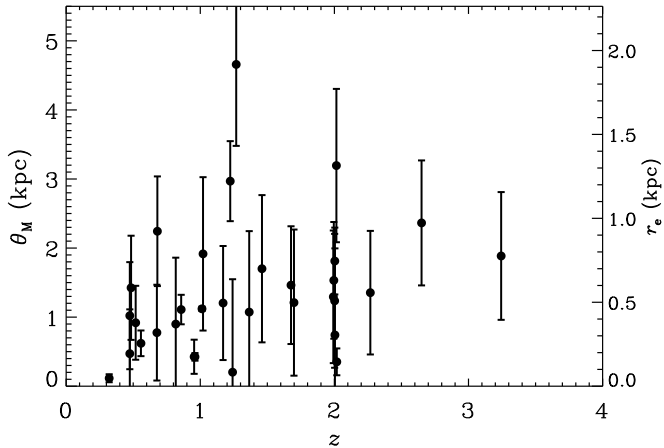}
\caption{The FWHM major-axis linear sizes of all 32 detected sources plotted against redshift.  The
  corresponding effective radii ($r_{\rm e}$) are marked along the
  right ordinate.  
  %Unresolved sources are marked with downward arrows.
  The FWHM linear sizes have a weighted median $ 1.3\pm0.28$\,kpc and a weighted rms
  scatter $\approx 0.79$\,kpc.  The corresponding effective radii
  $r_{\rm e}$ have a weighted median $509\pm114$\,pc and a weighted rms scatter $\approx
  324$\,pc.  No obvious evolution in 10\,GHz radio size is seen with
  redshift. }
\label{fig:dmaj_vs_z}
\end{figure}

\subsection{The Size Distribution of $z\sim1$ Star-Forming Disks}

%\subsection{Relating Radio and Optical Source Sizes \label{sec:optsiz}}

%XX JJC: Eric: You asked if I wanted to write a quick statement about 
%XX this (was Section 2.4, now Section 3.2), so I replaced
%XX the first paragraph by the next two paragraphs:
Radio astronomers 
traditionally fit elliptical Gaussians to sources on
images and specify source sizes by their deconvolved major and minor
FWHM axes $\theta_\mathrm{M}$ and $\theta_\mathrm{m}$.
However, the radio brightness distributions of star-forming spiral galaxies
are better approximated by optically thin exponential disks, and OIR
astronomers often characterize the size of a disk by the effective radius $r_{\rm
  e}$ that encloses half of the total flux density of the deprojected disk.  
Appendix~\ref{sec:AppendixB}
shows analytically that these two size measures are
related by
\begin{equation}
\theta_{\rm M} \approx 2.430\,r_{\rm e}~,\footnote{In the paper $r_{e}$ and $\theta_{\rm M}$ are used to describe both linear and angular sizes with appropriate units labeled.}
\end{equation}
which we use to compare our radio disk sizes to OIR sizes in the
literature.  

The major radio-astronomy image analysis packages (e.g., AIPS, CASA)
make Gaussian but not exponential fits.  These fitting routines are
not easily described analytically, so we performed numerical
simulations which show that Gaussian and exponential fits yield
comparable results for both deconvolved sizes and peak brightness-to-integrated
flux density ratios $S_\mathrm{P}/S_\mathrm{I}$
(Appendix~\ref{sec:AppendixC}), at least for sources with $r_\mathrm{e}
\lesssim \theta_{1/2}$.
%XX JJC: end of replacement paragraphs

%Of the 32 reliable 10\,GHz detections listed in Table \ref{tbl-1}, the major axes of 6  ($\approx 20$\%) are confidently resolved in the $0\,\farcs22$-resolution image.
The distribution of the deconvolved angular sizes for the 32 reliable 10\,GHz detections listed in Table \ref{tbl-1} is plotted in Figure \ref{fig:sizedist}.  
Among these, the major axes of 6 sources ($\approx 20$\%) are confidently resolved at $\geq 2\sigma_{\phi_{\rm M}}$ significance (see \S\ref{sec:srcfind}), and are identified in Table~\ref{tbl-1}.      
To determine the typical source size among the entire population of sources, we calculate a weighted median.  
For each source size, the weight is inversely proportional to the solid angle in which it could have been detected, limited either by $SNR$ or by the area within the 5\% primary beam cutoff.  
The constant of proportionality is set such that a source that could be detected anywhere within the survey area has weight $w_{\rm i} = 1$, but any constant of proportionality would give the same result.  
This weight is designed to exactly compensate for the fact that weak extended sources could be seen only in a fraction of the full survey field.  
Our weighted source size distribution should be representative of a sample that was uniformly selected, unbiased by angular size, over the entire survey area. 

The weighted median deconvolved source 
%XX JJC: For clarity, I replaced ``size'' by
FWHM
is $\langle \theta_{\mathrm M} \rangle = 167 \pm 32$\,mas, and the
weighted rms size scatter is $\approx 91$\,mas.  The corresponding weighted median
effective radius 
%(see \S\ref{sec:optsiz}) 
is $\langle r_{\mathrm e}
\rangle = 69\pm13$\,mas, and the weighted rms scatter in $r_\mathrm{e}$ is
$\approx 38 \mathrm{\,mas}$.  
%EJM:  I am not that bothered by it, since it is simply a difference in units, but not the actual measured quantity.  
%XX JJC: Our size notation in this section and Figure 8 is bad because
%XX it uses the same symbols to indicate both linear and angular sizes.
%XX Can we use something other than $\theta_\mathrm{M}$ to indicate 
%XX FWHM linear diameter; e.g. $D_{1/2}$?
%XX At the very least, we should explicitly say which we are talking about,
%XX as shown below.
Using the measured redshifts of these
resolved sources, we plotted their 
%XX JJC: inserted 'linear'
linear
sizes as a function of redshift in
Figure \ref{fig:dmaj_vs_z}.  
%plus 
%XX JJC: inserted 'linear'
%size upper limits for unresolved
%sources.  
The weighted median FWHM major axis 
%XX JJC: inserted 'linear'
linear
%XX JJC: replaced 'size' with
diameter
of these sources is $\langle
\theta_{\rm M} \rangle = 1.2\pm0.28$\,kpc, and $\theta_\mathrm{M}$ has a
weighted rms scatter of $\approx 0.79 \mathrm{\,kpc}$.  The
corresponding weighted median effective 
%XX JJC: inserted 'linear'
linear
radius is $\langle r_{\rm e} \rangle =
509\pm114$\,pc, and the rms scatter in $r_\mathrm{e}$ is $\approx 324
\mathrm{\,pc}$.  There is no obvious indication of 
%XX JJC: inserted 'linear', which is particularly needed here since we
%XX already speculated that the radio sources found in our low-resolution images
%XX have lower redshifts.
linear
size evolution with
redshift.

\subsubsection{Comparison with Other Radio and mm/sub-mm Sizes}
Using a deep VLA 3\,GHz image of the Lockman Hole made with $\theta_{1/2} =
0\,\farcs 66$ resolution, Condon et al. (2016, in prep) found a
preliminary median source size $\theta_\mathrm{M} \approx 300
\pm 100$\,mas among their detections.  Although slightly larger than
the 10\,GHz sizes reported here, they are actually quite consistent
since we expect the 3\,GHz radio size to be slightly larger than at
10\,GHz for two reasons: (1) The lower-energy relativistic electrons
emitting at 3\,GHz may diffuse farther because they have longer
synchrotron lifetimes.  
%For rigidity\footnote{Magnetic rigidity $R$ is defined as $R = pc/(Ze)$, where $p$ is momentum, $c$ is the speed of light, and $Ze$ is the particle charge.  Thus, for protons and electrons, $R = \sqrt{E_{\rm CR}^2 -  E_{0}^2}$), where $E_{\rm CR}$ is the CR energy and $E_{0}$ is the particle rest-mass energy.}-dependent diffusion of cosmic-rays (CRs) scaling as $R^{0.75}$, 
%XX jjc explain what is 'rigidity' R as measured for CR electrons and
For rigidity-dependent diffusion of cosmic-rays (CRs), where the magnetic rigidity\footnote{Magnetic rigidity $R$ is defined as $R = pc/(Ze)$, where $p$ is momentum, $c$ is the speed of light, and $Ze$ is the particle charge.  Thus, for protons and electrons, $R = \sqrt{E_{\rm CR}^2 -  E_{0}^2}$), where $E_{\rm CR}$ is the CR energy and $E_{0}$ is the particle rest-mass energy.} scales as $R^{0.75}$   
as measured for CR electrons and protons around 30\,Dor \citep{ejm12a}, 3\,GHz emitting CR electrons will diffuse $\approx 25$\% further than 10\,GHz emitting CR electrons. 
(2) Thermal emission confined to the
star-forming regions contributes a smaller fraction of the total flux
density at 3\,GHz.
%XX jjc notice that I added this comment about the thermal fraction
% ejm:  excellent point.  

However, the measured radio sizes of our 10\,GHz-selected sources are
significantly smaller than the 3\,GHz sizes reported by \citet{om17} 
for a sample of 115 known SMGs in the COSMOS field.  Using sensitive ($\sigma_\mathrm{n} \approx 2.4\,\mu
\mathrm{Jy\,beam}^{-1}$) 3\,GHz images from the VLA-COSMOS 3\,GHz
Large Project (Smol{\v c}i{\'c} et al. 2016, in press), they found a
%JJC: Should we change the radio sizes below from 590 mas and 420 mas to
% 0.59 arcsec and 0.42 arcsec for consistency with the radio size units
% elsewhere in this section?
median source size $\theta_\mathrm{M} \approx 590 \mathrm{\,mas}$ and
an rms size scatter of 420\,mas.  Given the redshifts of their
sources, their median angular size corresponds to a median linear size
$\approx 4.5 \mathrm{\,kpc}$. 
%XX JJC; Replaced 'it' by 'that' to make it clear we are comparing with 
%XX the Smolcic size:
That 
is also close to the
median 1.4\,GHz size for a sample of 12 SMGs reported by \citet[][i.e., $640\pm100$\,mas]{bi08}, which corresponds to 
%XX JJC: inserted 'a'
a
linear size of
$5$\,kpc given the redshifts of their sources.
%\citet[$0\,\farcs83\pm0\,\farcs5$][]{sc04} and \citet[$0\,\farcs51\pm0\,\farcs07$][]{bi08}, which correspond to $4-6$\,kpc%at the redshifts of these SMGs 
In contrast, our measured sizes are in
much better agreement with the ALMA sizes from \citet{si15} and
\citet{jm15}, who report median sizes of 
%JJC: Again, maybe use units of arcsec, not milliarcec here
$200\pm50$\,mas
($1.6\pm0.14$\,kpc) at 1.1\,mm and $300\pm40$\,mas ($2.4\pm0.2$\,kpc) at 870\,$\mu$m, respectively.  
%XX jjc at what frequency were the ALMA sizes measured? 
Thus, while our radio sizes are not for a sample of SMGs, which may be
different from the more typical star-forming galaxies 
selected at 10\,GHz, it is interesting to find that our radio sizes
are compatible with the high-resolution dust-emitting sizes of SMGs.

One might expect sub-mm sizes to be different from the radio emitting sizes given the different selection criteria.  
SMGs selected at sub-mm wavelengths may be intrinsically different from star-forming galaxies selected by synchrotron or free-free emission at cm wavelengths because the sub-mm galaxies have no $K$-correction and thus are intrinsically more luminous and at higher redshifts than cm-selected galaxies.    
Furthermore, even for the same galaxy population, the size measured at sub-mm wavelengths may be different (larger) than the size measured at cm wavelengths because the submm size is that of the cold cirrus dust distribution while the cm size is closer to that of the current star formation.

\begin{figure}
\epsscale{1.2}
\plotone{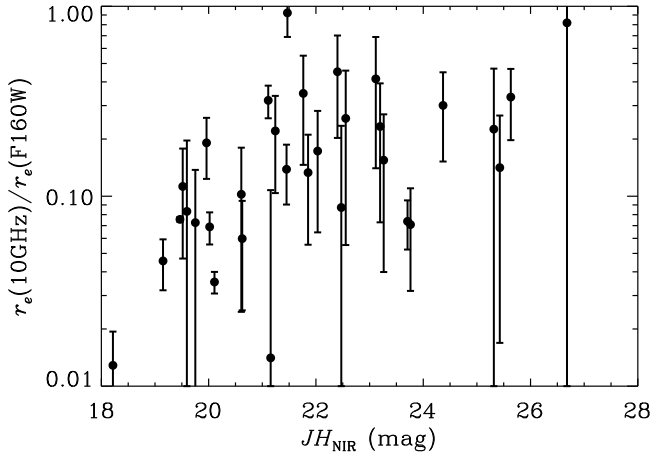}
\caption{ The ratio of 10\,GHz-to-{\it HST}/WFC3 
%XX JC inserted reference and 'continuum' to distinguish from Nelson et al. H\alpha
F160W  \citep{vdWel14} continuum
effective radii
  plotted as a function of NIR magnitude for all 32 sources reliably detected 
  in the full-resolution radio image.  
  %Sources only having an upper limit for their 10\,GHz size are indicated by a downward arrow.  
  On average, the 10\,GHz sizes are
  $0.14\pm0.05$ times the rest-frame optical size
%XX JJC inserted next line
(not corrected for extinction), 
and the size ratios
  have an rms scatter of 0.21, indicating that the star formation in
  our sample of $z\sim1$ star-forming galaxies is centrally
  concentrated.  }
\label{fig:szratmag}
\end{figure}
%JJC: Does the point at 18 mag with radio/optical size
% ratio only 0.014 make sense?  Is that an especially large galaxy whose radio size
% we may have underestimated, or might it be an AGN whose radio size does not
% reflect size of the star-forming region?

\begin{figure*}
\epsscale{1.}
\plottwo{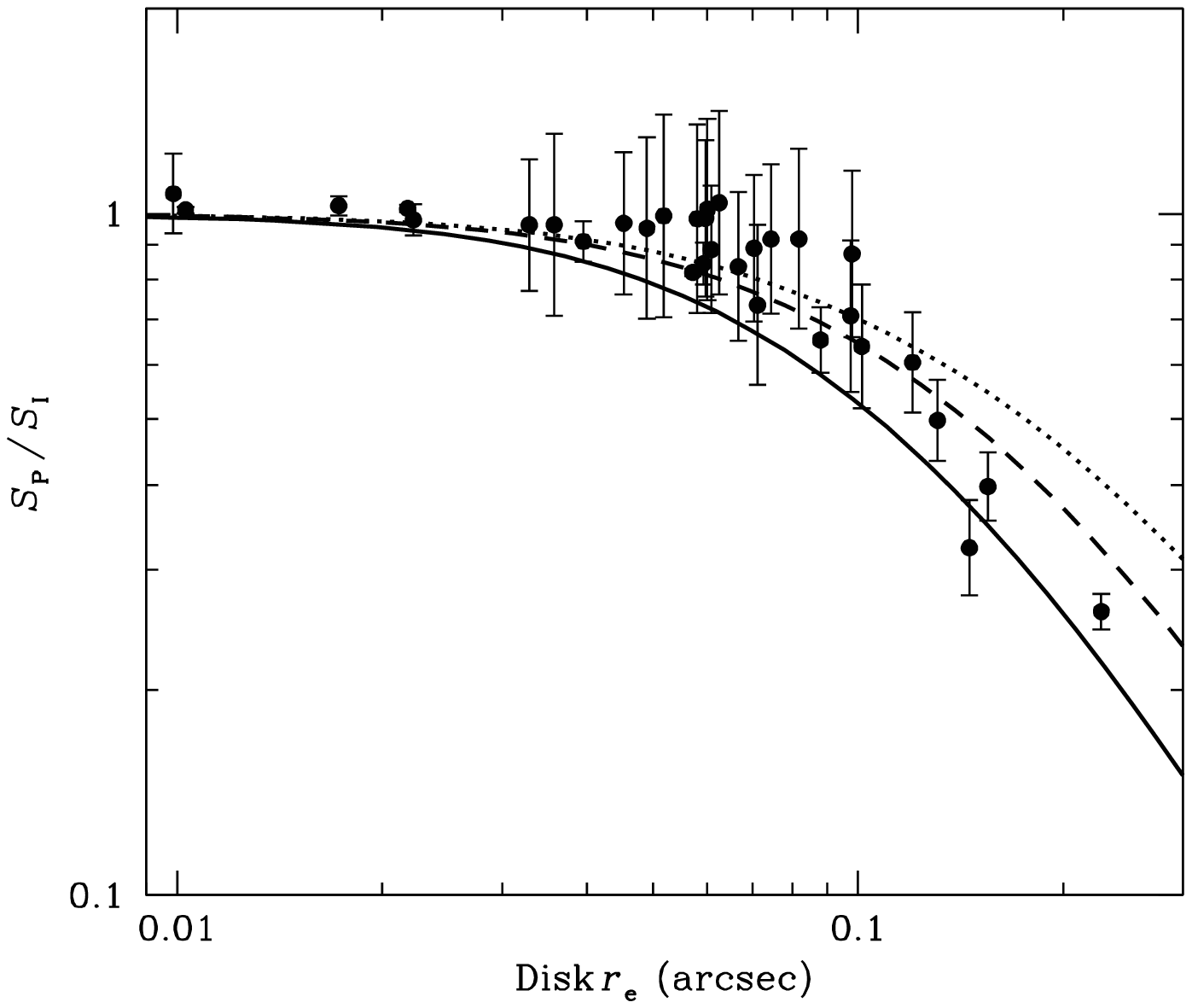}{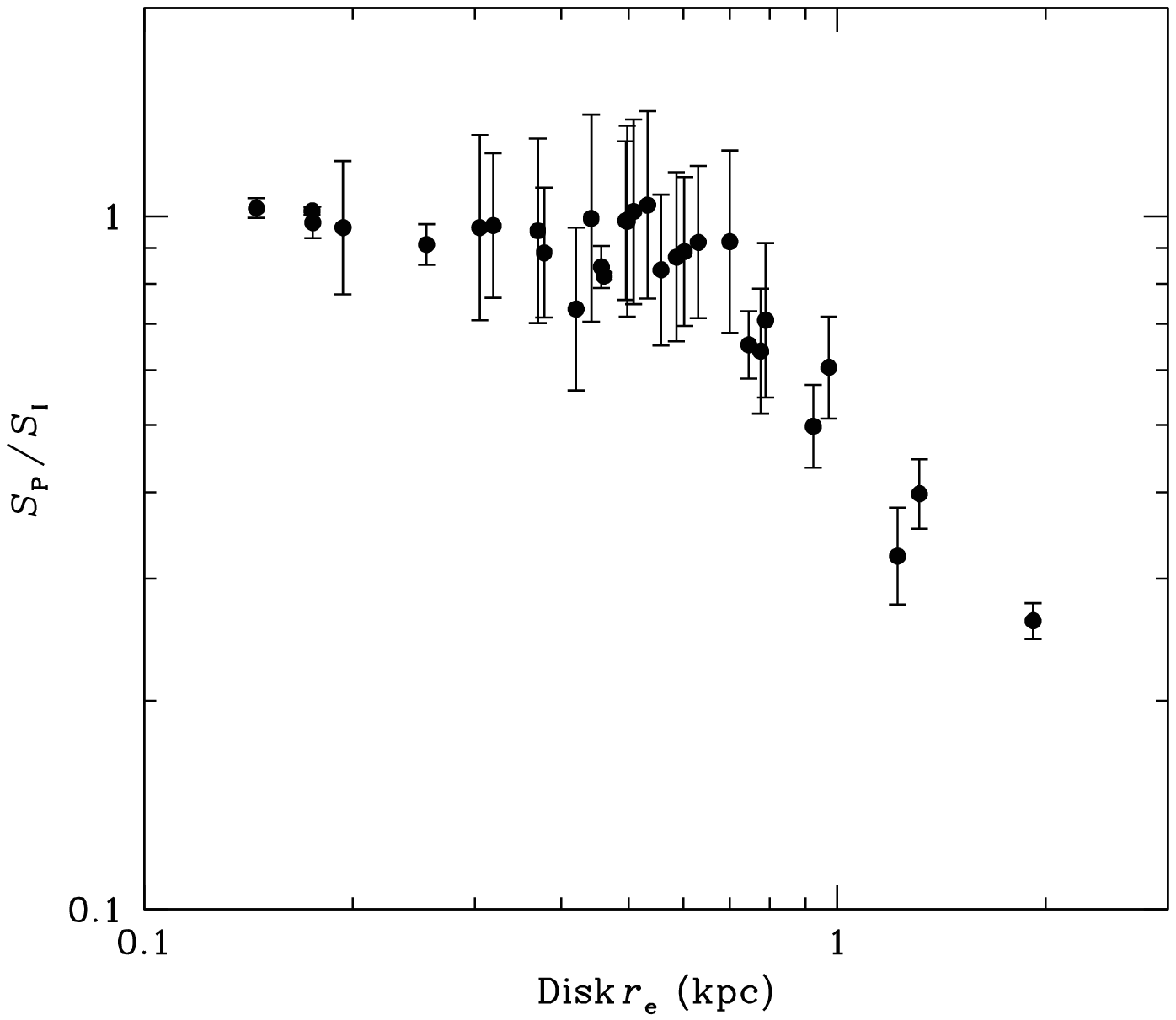}
\caption{{\it Left:} Measured ratios $S_\mathrm{P} /
  S_\mathrm{I}$ of peak brightness-to-integrated flux densities for all sources
  detected in our $0\,\farcs22$-resolution image 
%   both resolved (filled circles) or unresolved (open circles), 
  versus the deconvolved disk effective radius $r_\mathrm{e}$ in units of arcsec.  
  %The open circles (unresolved sources) mark upper limits to $r_\mathrm{e}$, so better data will move them to the left.  
  The solid curve was calculated from
  Equation~\ref{eqn:spsi} for a circular (face-on) exponential disk.
  The dashed curve was calculated from Equation~\ref{eqn:spsiapprox}
  for the median inclination angle $i = 60^\circ$ [$\cos(i) = 0.5$] of
  randomly oriented disks, and the dotted curve corresponds to a
  nearly edge-on disk with $\cos(i) = 0.25$. The data points should lie between the solid and dotted curves and be centered around the
dashed curve. 
%The open circles should lie above all of the curves.
{\it Right:} The same as the left panel, except that the abscissa units are kpc instead of arcsec for sources with measured redshifts.}
\label{fig:p2ivsre}
\end{figure*}

\subsubsection{Comparing Radio, H$\alpha$, and Optical Sizes}

We also compared our star-forming galaxy disk sizes with those
reported by \citet{ejn16}, which are based on stacking resolved
H$\alpha$ and H$\beta$ emission-line images from the 3D-HST survey. By
comparing their H$\alpha$ and H$\beta$ galaxy images, these authors
could apply an extinction correction to their H$\alpha$ images before
fitting for the effective radius.  This extinction correction
significantly lowers the measured sizes because it multiplies the
inferred star-formation rates in the central $r < 1$\,kpc of
galaxies with $9.8 < \log(M/M_{\odot}) < 11.0$ by a factor of
$\approx 6$.  The average extinction-corrected H$\alpha$ radial profile
of these galaxies declines by a factor of $\approx 100$ between the
center and $r \approx 2$\,kpc, which corresponds to an effective
radius of $\langle r_{\mathrm e}({\mathrm H}\alpha)\rangle \approx
0.73$\,kpc.  Before making the extinction correction, \citet{ejn12}
had reported a median H$\alpha$ effective radius $\langle r_{\mathrm
  e} \rangle = 4.2\pm0.1$\,kpc for a sample of 57 strongly
star-forming galaxies in the same mass range. Thus, the uncorrected
$r_\mathrm{e}$ is $\approx 6 \times$ larger than the corrected
$r_\mathrm{e}$.  The effective radius of the extinction-corrected
H$\alpha$ radial profiles is not significantly larger than our
weighted median $\langle r_\mathrm{e}\rangle \approx 0.51\pm0.11 \mathrm{\,kpc}$ with rms scatter 0.32\,kpc at 10\,GHz and is much more
consistent with the radio sizes reported here than with others in the
literature.  The small difference between our measured sizes and the
extinction-corrected H$\alpha$ sizes might result from underestimating
extinctions in the centers of the most massive galaxies in
\citet{ejn16}.

The effect of extinction may also be contributing to what is found in Figure \ref{fig:szratmag}, which plots the ratio of the 10\,GHz effective
radius to the rest-frame optical continuum $r_\mathrm{e}$ reported by \citet{vdWel14} as a function of NIR magnitude.  
The rest-frame optical data used were not corrected for extinction.  
The star-forming disks measured at 10\,GHz, which are similar to the extinction-corrected H$\alpha$ sizes reported by \citet{ejn16}, are significantly smaller than the
uncorrected rest-frame optical continuum sizes.  
The median 10\,GHz-to-F160W size ratio is $0.14\pm0.05$, and the size ratios have an rms scatter of 0.21.  
Consequently, the star formation appears to be centrally concentrated in this sample of $z\sim1$ galaxies.
%However, similar to what has been shown for H$\alpha$ sizes, extinction may also be playing a role.
%jjc We should distinguish the optical size measured in Halpha from the optical
%jjc size measured in the continuum because the Halpha size is the size of the
%jjc region where new massive stars are forming, while the optical continuum
%jjc size is the size includes older stars.  Also, the amount of extinction
%jjc in the dusty molecular clouds were massive ionizing stars are currently
%jjc found is likely to be greater than the amount of extinction suffered
%jjc where old stars are found, just as in our own Galaxy

Here, we have converted the decononvolved Gaussian fits to the radio sizes to half-light radii $r_e$ assuming the conversion for exponential profiles discussed in Appendix~\ref{sec:AppendixB}.  
The {\it HST}/WFC3 sizes from \citet{vdWel14}, $r_{\rm e}{\rm(F160W)}$, were derived from \citet{jls63,jls68} profile model fitting, and thus may not exactly match exponential fits (i.e., S\'{e}rsic index $n=1$) on a galaxy-by-galaxy basis.  
Moreover, the {\it HST}/WFC3 F160W filter measures optical rest-frame starlight at the mean redshift of our 10\,GHz-selected sample, and thus may not simply reflect the distribution of optically-bright star formation in each galaxy.   
Nevertheless, it seems clear from Figure \ref{fig:szratmag} that the 10\,GHz radio (and H$\alpha$) sizes are uniformly and significantly smaller than the optical galaxy sizes derived from the CANDELS {\it HST} 
%XX JJC: replaced 'imaging' by
images not corrected for extinction.
Accordingly, similar to what has been shown for H$\alpha$ sizes, extinction may be playing a role.

%XX JJC: Eric: Here you asked me to ``please respond'' to the referee's comments about
%XX van der Wel's size evolution as a function of stellar mass.  I presume that the
%XX galaxies that we were able to detect at 10 GHz are all fairly massive galaxies,
%XX while the HST image used by van der Wel can detect very faint low-mass galaxies.
%XX If that is the case, maybe our galaxies don't sample the low-mass end well enough
%XX for us to say anything significant about size evolution as a function of 
%XX stellar mass.

%Using the NIR continuum magnitude as a crude proxy for stellar mass of the
%galaxy, we see that the ratio of the 10\,GHz-to-optical size is
%roughly independent of stellar mass, perhaps showing a weak trend of
%increasing with decreasing stellar mass.  As shown in \citet{ejn16},
%the measured extinction of $z\sim1$ star-forming galaxies appears to
%increase with stellar mass.  
%Thus, the weak trend in Figure \ref{fig:szratmag} might simply result from the rest-frame optical
%continuum sizes being overestimated with increasing stellar mass and
%obscuration, suggesting that uncorrected optical continuum sizes should also be
%interpreted with caution.

\subsubsection{$S_\mathrm{P} / S_\mathrm{I}$ Ratios of Randomly Oriented Thin Disks}

If star-forming galaxies are randomly oriented, transparent, thin, and
circular exponential disks at radio wavelengths, they should appear
elliptical on the sky, with minor axes shortened by $
\theta_\mathrm{m} / \theta_\mathrm{M} = \cos(i)$, where $i$ is the
inclination angle between the disk normal and the line of sight.
However, the minor axes $\theta_\mathrm{m}$ of many galaxies in
Table~\ref{tbl-1} are not resolved, which necessitates another way to check
this hypothesis.  One way is to use the ratio $S_{\rm P} / S_{\rm I}$ of peak
  brightness to integrated flux density, a quantity that depends on
  both $\theta_\mathrm{M}$ and $\theta_\mathrm{m}$, and is available
  for all galaxies.  The dependence of $S_\mathrm{P} / S_\mathrm{I}$
  on galaxy size and inclination is derived in Appendix~\ref{sec:AppendixC}. The results
  are shown as functions of $r_\mathrm{e}$ in Figure~\ref{fig:p2ivsre}.
  They are consistent with most of our radio sources being randomly
  oriented, thin, circular exponential disks.

%% ASK JIM
% I speculate that the 0.22 arcsec image discriminates against 0.3 arcsec FWHM
% radio sources.  We need to calculate a correction for sources missing because
% their peak flux densities are < 5 sigma even though their total flux 
% densities are > 5 sigma.  This correction is similar to the resolution
% correction used for source counts.

%It is worth noting that while we have limited source characterization to those detected down to $\approx$5\% of the primary beam response;  
%, an additional 5 source components were detected outside of this region, two of which could be matched to an OIR counterpart with a measured redshift  optical counterparts having redshifts.  
%The J2000 coordinate and redshifts of these sources are: 
%$\alpha=12^\mathrm{h}36^\mathrm{m}23\fs54, \delta = +62\degr16\arcmin42\farcs7$, $z=1.918$; 
%$\alpha=12^\mathrm{h}36^\mathrm{m}59\fs33, \delta = +62\degr18\arcmin32\farcs6$, $z=1.995$.  
%Consequently, by having the full survey data, we will be able to robustly characterize these sources which will be covered by multiple pointings.  

\begin{figure}
\epsscale{1.2}
\plotone{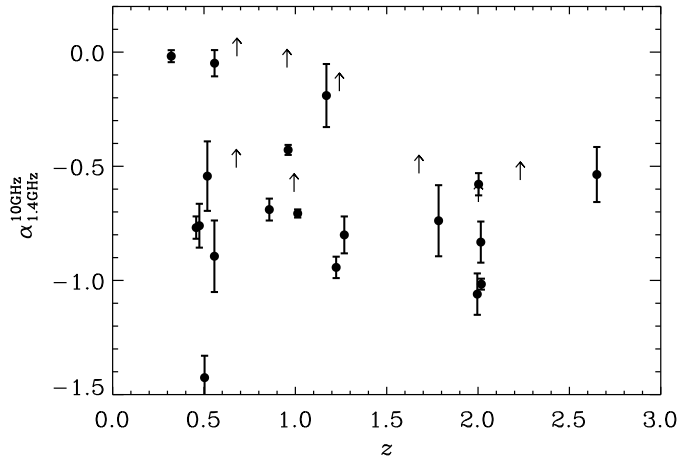}
\caption{
The median spectral index among all 19, 1.4\,GHz detected sources is 
$\langle \alpha_{\rm 1.4\,GHz}^{\rm 10\,GHz} \rangle \approx -0.74\pm0.10$ with a standard deviation of 0.35.
%% CALCULATE
%XX I presume that standard deviation is the rms scatter in alpha, not the smaller rms
%XX uncertainty in the median.  
Including the additional 8 sources for which there are only upper limits on the 1.4\,GHz flux density flattens the median spectral index 
%XX I inserted the next line: 
of our full flux-limited sample selected at 10 GHz to 
$\langle \alpha_{\rm 1.4\,GHz}^{\rm 10\,GHz} \rangle \gtrsim -0.61$.}
\label{fig:spx_vs_z}
\end{figure}

%jjc Shouldn't this be merged with Section 2.5?
\subsection{Spectral Indices and Thermal Fractions}
%XX rephrasing below
Using the 1\arcsec~and 2\arcsec~resolution images, which 
bracket the $1\,\farcs7$ resolution of the \citet{gm10} 1.4\,GHz image, we are able to measure radio spectral indices with nearly matched resolution 
(see Table \ref{tbl-2}).  
Among the 19 sources with 1.4\,GHz counterparts, the median observed spectral index between 1.4 and 10\,GHz is $\langle \alpha_{\rm 1.4\,GHz}^{\rm 10\,GHz} \rangle \approx -0.74\pm0.10$ with a standard deviation of $\sigma_{\alpha_{\rm 1.4\,GHz}^{\rm 10\,GHz}} \approx 0.35$.  
%%XX CALCULATE
%XX scatter in alpha or in median alpha?  I replaced 'value' with 'median'
%XX below because the 0.36 scatter in alpha is greater than observed locally,
%XX most likely because the errors in alpha are larger.  We should put
%XX in error estimates for alpha.  
%XX If fluxes S_1 +- Sigma_1 and S_2 +- Sigma_2 are measured at frequencies
%XX f_1 and f_2 respectively, then the rms error in alpha is
%XX sigma_alpha = [(Sigma_1 / S_1)^2 + (Sigma_2 / S_2)^2]^{1/2} / |ln (S_1 / S_2)|
%XX e.g., if SNR = 5 at both f_1 = 10 and f_2 = 1.4, then sigma_alpha = 0.14
This median is consistent with what is measured for star-forming galaxies in the local universe \citep[e.g., ][]{jc92}.  
However, by including the 8 sources (30\% of our detections) having only 1.4\,GHz upper limits, the median flattens significantly to $\gtrsim-0.61$.  
The large number of sources without 1.4\,GHz counterparts is not that surprising since, as stated in \S\ref{sec:imging}, for a source with spectral index of $-0.7$ the 10\,GHz images are $\approx 2$ times more sensitive than the 1.4\,GHz image of \citet{gm10}.  
Even so, it appears that there is a significant fraction of sources at $z\gtrsim1$ having somewhat flat spectral indices, suggesting that higher-frequency radio measurements may indeed be more sensitive to free-free emission and consequently a more robust measure for the current star formation activity in such systems.  
Figure \ref{fig:spx_vs_z} plots the measured spectral indices agains redshift, for which there is no clear trend.  

As discussed in \S\ref{sec:spx}, by assuming a fixed, non-thermal spectral index for each source, we can use the measured spectral indices to estimate the fractional contributions from thermal emission at the rest-frame frequency $(\nu_{\rm r}/{\rm GHz}) = 10(1+z)$ \citep[e.g., ][]{kwb84,ejm12b}, which are listed in Table \ref{tbl-2}. 
%% TALK TO JIM
% How much do we know about the synchrotron spectral index at ~ 20 GHz in the
% source frame?  Maybe it is steeper than 0.85?
%XX JJC: We don't know much about the synchrotron spectral index at  ~20 GHz an up in 
%XX the source frame because accurate high-frequency total flux density measurements
%XX of extended, low brightness sources are difficult to make, even with single
%XX dishes.  Mainly we know from theory (and from observations of AGN-powered radio
%XX galaxies) that synchrotron spectra do steepen at high frequencies owing to
%XX synchrotron and inverse-Compton losses.  Thus the referee said ``the caveat that is
%XX mentioned with respect to possible spectral steepening at 20 GHz rest-frame is a
%XX bit unconclusive [sic]''.  Our caveat SHOULD be a bit inconclusive because 
%XX our knowledge is a bit inconclusive.
These estimates additionally assume that the observed 10\,GHz emission is in fact powered by star formation.  
For the 19 sources having 1.4\,GHz counterparts the median redshift corresponds to a rest-frame frequency of $\approx 20$\,GHz for which we estimate a median thermal fraction of $26\pm0.09$\% with an rms scatter of 31\%.  
If we additionally include the 8 sources for which we only have upper limits at 1.4\,GHz, the median redshift also corresponds to a rest-frame frequency of $\approx 20$\,GHz and places a lower limit on the median thermal fraction of $\gtrsim$48\%.  
While we have assumed a non-thermal spectral index of $-0.85$, one caveat is that spectral steepening in the rest-frame near 20\,GHz (e.g., from increased synchrotron and inverse-Compton losses), which is extremely difficult to measure, could result in underestimates of the thermal fractions.

\subsubsection{Comparison with Previous Deep X-band Imaging}

Our 2\arcsec~tapered image can be compared with the X-band image of
\citet{richards98}.  They detected 29 sources stronger than
5$\sigma_\mathrm{n}$ in a field of radius of 4\farcm6 (truncated at
$\approx 8$\% of the primary beam response) in their 8.5\,GHz image
centered on the {\it Hubble} Deep Field.  For the original VLA system, the
primary beam FWHM at 8.5\,GHz is $\approx 5\farcm2$.  Their 8.5\,GHz VLA data from
the A, BnA, C, DnC, and D configurations respond to sources up to
10\arcsec~in size.  The bulk of their observing time was in the C
configuration, and their final image had resolution $\theta_{1/2}
\approx 3\farcs5$ and rms noise $\sigma_\mathrm{n} \approx
1.8\,\mu\mathrm{Jy\,beam}^{-1}$, corresponding to a
brightness-temperature sensitivity $\sigma_\mathrm{n} \approx 2.5
\,\mathrm{\,mK}$.  This is similar to our $\sigma_\mathrm{n} = 1.5\,\mu
\mathrm{Jy\,beam}^{-1}$ rms, which scales to $\approx
1.7\,\mu\mathrm{Jy\,beam}^{-1}$ at 8.5\,GHz for a source with typical
spectral index $\alpha = -0.7$, and a 4.7\,mK brightness temperature
rms achieved in our 2\arcsec~image.

In our 2\arcsec~resolution image we detected a total of 14 sources
with $S_\mathrm{P}/\sigma_\mathrm{n} \geq 5$ that are reliable (i.e., the
same 5$\sigma_\mathrm{n}$ sources we consider to be reliable in our
1\arcsec-resolution image).  Given the ratio of primary beam solid
angles $\Omega_{\mathrm{10 GHz}}/ \Omega_{\mathrm{8.5 GHz}} = 0.67$
and sensitivities at the two frequencies, we would expect them to detect
$\approx 50$\% more sources (i.e., $\approx 21$ sources).  
%XX jjc What is ``this''?  I am confused by the sentence below.
% ejm: this = the above estimate for how many sources we expect them to detect
This prediction is slightly more than 1$\sigma$ smaller than what counting errors would
suggest.  However, if we were to include the single 5$\sigma$
detection from our 2\arcsec~tapered image that is considered unreliable,
or exclude the four $S_\mathrm{P} > 5\sigma_\mathrm{n}$ sources in the
\citet{richards98} sample that were not matched to optical
counterparts, our numbers do in fact agree to within 1$\sigma$ of the
counting errors.

\subsection{A Revised Redshift and Spectral Energy Distribution Analysis for VLA~J123642$+$621331}

We detected the radio source VLA J123642$+$621331 described by
\citet{waddington99} as a compact [$r_{\rm e}{\rm(F160W)} \approx 0\,\farcs2$]
%XX jjc Why do you call r_e = 0.2 arcsec ``small''?  It is bigger
%XX jjc than most of the galaxies in the left panel of Figure 9.  dusty
star-forming galaxy containing an AGN.  The 10\,GHz
flux density measured in our 1\arcsec~ tapered image is consistent
with the 8.5\,GHz flux density reported by
\citet{richards98,richards99}, being $ \approx 70\,\mu$Jy.
The source is detected in both our $0\,\farcs 22$ and $1
\arcsec$ resolution images, having deconvolved major axes of
$42\pm23$\,mas and $0\,\farcs65\pm0\,\farcs06$, respectively, suggesting a
compact core and consistent with the upper limit $ < 0\,\farcs1$
reported by \citet{richards98} in their lower signal-to-noise
0\,\farcs2 resolution image.  VLA J123642$+$621331 was resolved by
high resolution (0\,\farcs15) 1.4\,GHz VLA+MERLIN observations
\citep{twm99}, which showed that 10\% of the flux density resides in
an extended component lying to the east of an unresolved core.  The
unresolved, compact core was additionally detected by the European
VLBI Network (EVN) at 1.6\,GHz, providing an upper limit on the core
size of 20\,mas and corresponding 1.6\,GHz brightness temperature of
$T_{\rm b} > 2\times10^{5}$\,K \citep{garrett01}, indicating that the
core emission is dominated by an AGN \citep{jc91}.  
%XX jjc Cite Condon, J. J., Huang, Z.-P., Yin, Q. F., & Thuan, T. X. T.
%XX jjc 1991, ApJ, 378, 65 as a reference for the claim that Tb > 2e5 K
%XX jjc is too bright for a starburst.
%However, as shown in Figure \ref{fit:wad-spec}. the combination of
%{\it HST}/WFC3 G140 grism and MOSFIRE spectroscopy indicate that the
%redshift of this source is almost certainly $z = 2.018\pm0.03$ and not the
%$z=4.424$ reported by \citet{waddington99} on the basis of a single
%line detected at 6595\,\AA~that was attributed to Ly$\alpha$.

\begin{figure}
\begin{center}
\epsscale{1.35}
\hspace*{-24pt}
\plotone{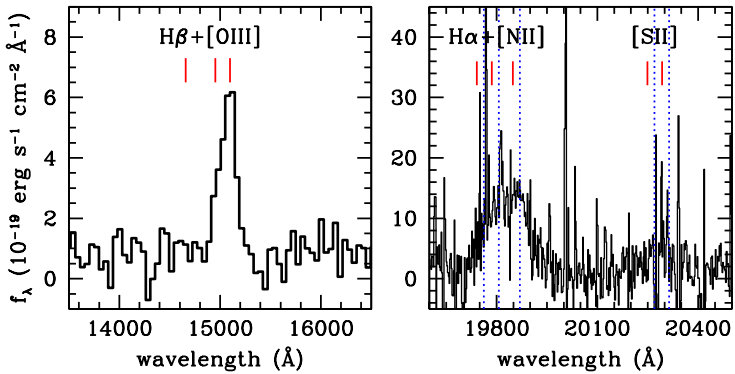}
\caption{
{\it Left:} A portion of the {\it HST}/WFC3 G140 grism spectrum of VLA~J123642+621331, showing the emission feature identified as the [O{\sc iii}] doublet.  
{\it Right:}  A portion of the Keck/MOSFIRE spectrum.  
The red tickmarks indicate the predicted wavelengths of the [N{\sc ii}] and H$\alpha$ emission lines, based on the redshift fit to the WFC3 grism spectrum ($z_{\rm [O\textsc{iii}]} = 2.015)$.  
The broad, blended H$\alpha$+[N{\sc ii}] complex seems to be redshifted slightly with respect to these predictions, with a visual estimate ($z = 2.018\pm0.003$) indicated by the dashed blue lines.
%The combined {\it HST}/WFC3 G140 grism and MOSFIRE spectra for VLA J123642$+$621331.  Identified lines (i.e., [OIII]\,5007\,\AA, H$\alpha$+[NII]) are shown, which provide a formal redshfit fit of $z=2.018\pm0.03$, which is in stark contrast with the originally claimed redshift of $z=4.424$ by \citet{waddington99} on the basis of a single line detected at 6595\,\AA~that was attributed to Ly$\alpha$.}
}
\label{fig:wad-spec}
\end{center}
\end{figure}

\citet{waddington99} reported a redshift $z = 4.424$ for this source, on the basis of a single emission line detected with Keck/LRIS at 6595\,\AA\ interpreted to be Lyman~$\alpha$, apparently offset by about $1^{\prime\prime}$ from the optical galaxy position. 
However, more recent multiwavelength photometric data in the GOODS-N field suggest that the galaxy may instead have a lower redshift, due to faint but significant detection in the GOODS {\it HST}/ACS $B$-band (F435W) image, as well as mid- to far-infrared photometry (and sub-mm upper limits) that seem inconsistent with $z = 4.4$.  
A slitless spectrum from the {\it HST}/WFC3 G140 grism (GO-11600, PI Benjamin Weiner) detects a strong, slightly asymmetric line (Figure \ref{fig:wad-spec}, left), which we interpret as the blended [O{\sc iii}] 4959,5007\,\AA\ doublet, with no detectable H$\beta$ [$f(5007)/f({\rm H}\beta) > 4.6$ at $2\sigma$].  
A constrained two-Gaussian fit to the extracted grism spectrum yields a redshift $z_{\rm [O\textsc{iii}]} = 2.015$.  

We then obtained a $K$-band spectrum of VLA~J123642$+$621331 with MOSFIRE on the Keck~1 telescope on UT 2014 April 13 under photometric conditions, using a slit width of 0\,\farcs7, for a total exposure time of 84 minutes.  
The data were reduced using the standard MOSFIRE data reduction pipeline (version 2014.06.10).  
Figure~\ref{fig:wad-spec} (right) shows a 2\arcsec~wide extraction for a portion of the spectrum. 
A broad ($\approx 175$\,\AA) feature is detected, centered at approximately 19825\,\AA.  
We interpret this as a blend of broad H$\alpha$ plus [N{\sc ii}].  
The complex appears to be slightly offset from the wavelengths predicted based on the fit to [O{\sc iii}] in the grism data.  
We cannot formally fit the lines, but we estimate $z = 2.018$ from the MOSFIRE spectrum.  

This redshift difference with respect to that from [O{\sc iii}] ($\Delta z = 0.003$) is easily consistent with typical uncertainties in {\it HST}/WFC3 grism spectral measurements \citep{im16}.    
The 6595\,\AA\ line reported by \citet{waddington99} would not correspond to any emission features commonly seen in distant galaxies or AGN, and it may be a serendipitous detection of another faint, nearby galaxy at a different redshift (perhaps indeed Lyman~$\alpha$), or it could be spurious.  
The apparently broad H$\alpha$ emission, high [O{\sc iii}]/H$\beta$ ratio, and perhaps strong [N{\sc ii}] all suggest that an AGN dominates the optical rest-frame nebular line emission.  
This IR-luminous, radio-loud AGN could be another member of an over-dense structure at $\langle z \rangle = 1.99$ traced by sub-mm-, radio-, and UV-selected star-forming galaxies \citep{sc09}.  

The WFC3 grism spectrum of J123642$+$621331 was also analyzed by \citet{rc14}, who derived $z = 2.018$, and in the 3D-HST catalog of \citet{im16}, who derived $z = 2.012 \pm 0.002$ (68\% confidence). 

\begin{figure}
\epsscale{1.2}
\plotone{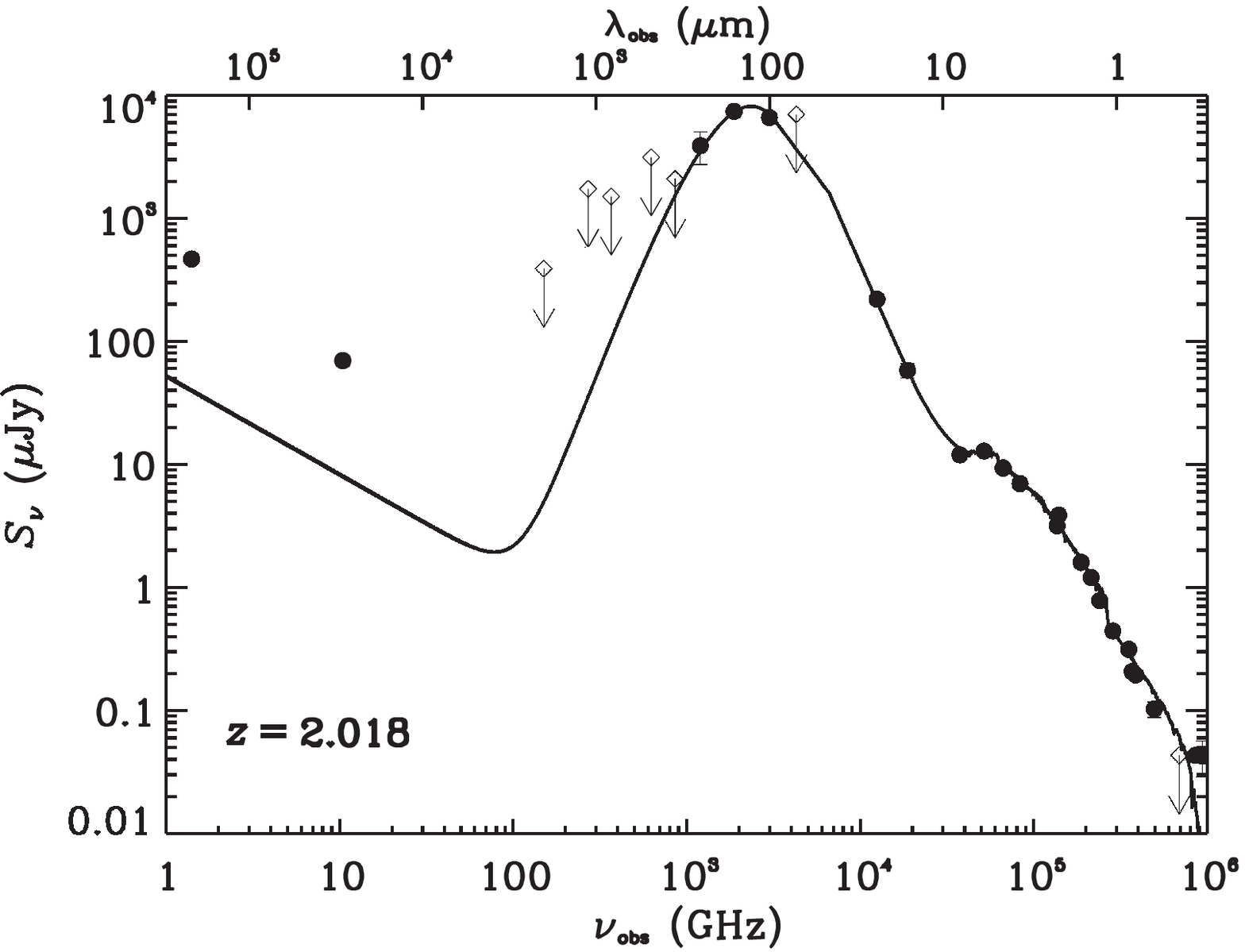}
\caption{
The best-fit radio-to-optical spectral energy distribution of VLA~J123642$+$621331.  
Data used in the fitting was taken from the literature: OIR data -- CANDLES GOODS-N multiwavelength catalog (G. Barro et al., 2016, in preparation); {\it Spitzer} and {\it Herschel} far-infrared data 
%\citep[][H. Inami et al., 2016, in preparation]{bm11,ht11}; 
(\citealt{bm11,ht11}; H. Inami et al., 2016, in preparation);
SCUBA sub-mm data \citep{cb03,ap05}; 1.16\,mm AzTEC$+$MAMBO data \citep{kp11}; 2mm GISMO data \citep{jgs14}.  
The radio data were not used to constrain the fit, which assumes the standard far-infrared--radio correlation and a typical radio spectrum with $S_{\nu} \propto \nu^{-0.8}$.   
The best-fit spectral energy distribution is characterized by  a stellar mass of $M_{*} = 2.4\times10^{10}\,M_{\odot}$, a stellar mass fraction of $f_{*} = 0.4$, an infrared (IR; $8-1000\,\mu$m) luminosity of $L_{\rm IR} = 2.3\times10^{12}\,L_{\odot}$, and a dust temperature of $T_{\rm dust} = 70\,$K, which is extremely hot even when compared to local AGN \citep[e.g.,][]{ls95,pa02}.    
The measured logarithmic far-infrared--radio ratio of $q=1.19$ has a factor of $\approx 30$ times (i.e., 5.6$\sigma$) more radio emission compared to the locally measured value for star-forming galaxies, also indicating the presence of an AGN.  
}
\label{fig:waddsed}
\end{figure}

Using a compilation of radio-to-optical data in the literature, we fit the full spectral energy distribution of VLA J123642$+$621331, which is shown in Figure \ref{fig:waddsed}.  
The radio data were not used to constrain the fit, which assumes the standard far-infrared--radio correlation and a typical radio spectrum with $S_{\nu} \propto \nu^{-0.8}$.  
The OIR data were fit with the updated \citet{bc03} stellar templates 
%, which include a revised prescription for stars on the asymptotic giant branch \citep[see][]{gb07},  
having an exponentially declining star formation history with a characteristic timescale of $\tau=0.1$\,Gyr and being extincted by an $A_{V} = 2.4$ assuming a local starburst attenuation law \citep{dc00}.  
The mid-infrared emission powered by hot dust was fit by a power law, while the far-infrared is fit by a cold dust model (i.e., a modified black body with $\beta_{\rm dust} = 1.9$).  
The best-fit spectral energy distribution is characterized by  a stellar mass of $M_{*} = 2.4\times10^{10}\,M_{\odot}$, a stellar mass fraction of $f_{*} = 0.4$, an IR luminosity of $L_{\rm IR} = 2.3\times10^{12}\,L_{\odot}$, and a dust temperature of $T_{\rm dust} = 70\,$K.  

Taking the observed 1.4\,GHz flux density of 494.2\,$\mu$Jy \citep{gm10} and the measured 1.4-to-10\,GHz spectral index of $-1.02$ in Table \ref{tbl-2}, the corresponding K-corrected logarithmic IR--radio ratio is $q_{\rm IR}=1.19$, which exhibits a factor of $\approx 30$ (i.e., 5.6$\sigma$) radio excess compared to the locally measured value of 2.64 \citep{efb03}, and thus strongly suggests the presence of an AGN.  
The far-infrared emission peak at 30$\,\mu$m in the rest frame implies a remarkably high dust temperature, similar to that seen in a minority of 3C sources \citep[e.g.,][]{ls95,pa02}, and suggests that the AGN dominates the bolometric luminosity of this source.  
%What is also remarkable about this source is the extremely high dust temperature, having its far-infrared emission peak at $\approx$30\,$\mu$m in the rest frame, similar to a minority of 3C sources.  

We have presented exquisite multiwavelength data on a radio galaxy at $z=2.018$.  
The change in redshift for this source from the previously claimed $z=4.424$ highlights the dangers in single line redshift determination, especially when the ancillary photometry is poor; for example, similar criteria have been used to claim the detection of a $z=4.88$ radio galaxy \citep{mjj09}.
With the accurate spectral energy distribution  we have of VLA~J123642$+$621331, we can predict the colors of $z>2$ radio galaxies that will be detected in forthcoming wide-area radio surveys such as the Evolutionary Map of the Universe \citep[EMU;][1.4\,GHz; 5$\sigma \approx 50\,\mu$Jy; $\theta_{1/2} \approx 10\arcsec$]{emu11} and 
VLA Sky Survey \citep[VLASS;][3\,GHz; 5$\sigma \approx 345\,\mu$Jy; $\theta_{1/2} \approx 2.5\arcsec$]{vlass15}.  
%EMU (Norris et al. 2011; 1.4 GHz; 5 $\sigma=50\,\mu$Jy; 10$\arcsec$ FWHM)  
%and VLASS (3 GHz; 5$\sigma$=350$\mu$Jy; 2.5$\arcsec$ FWHM).
Sources like VLA~J123642$+$621331 at $z\sim3$ would have flux densities of 0.45\,$\mu$Jy in the $H-$band, 0.9\,$\mu$Jy in the $K-$band and $S_{\rm 1.4\,GHz}$/$S_{K}$ ratios of $\approx 170$. 
It would therefore be challenging to detect their counterparts in wide-field NIR surveys such as those that will be undertaken with {\it EUCLID}. 
At $z\sim5$, which is the limit at which such a source would be detected in the EMU survey, it would be 60\,nJy in the $H-$band, 130\,nJy in the $K-$band and 0.5\,$\mu$Jy at 3.6\,$\mu$m. 
These are challenging sensitivities to achieve, partly due to source confusion in the latter case.  
Confirming the redshifts of such sources will instead benefit from spectroscopy at sub-mm/mm frequencies since the equivalent $z\sim5$ source would be 150\,$\mu$Jy and 80\,$\mu$Jy at 850\,$\mu$m and 1.2\,mm respectively, which is achievable with ALMA and NOEMA. 
In summary, the low stellar mass of VLA~J123642Q$+$621331 suggests that it will be difficult to obtain OIR spectroscopic redshifts of high-redshift ($z>3$) radio galaxy candidates and mm spectroscopy of [C{\sc ii}]  may be the most compelling approach.

\section{Conclusions}

In this paper we presented results from our pilot VLA 10\,GHz survey of
GOODS-N, aimed at resolving compact starbursts in the redshift range
between $1 \lesssim z \lesssim 3$.  This deep, single pointing image
reaches an rms noise $\sigma_\mathrm{n} = 572\mathrm{\,nJy\,beam}^{-1}$ and has $\theta_{1/2} = 0\,\farcs22$ resolution.  Our conclusions can be summarized as follows:

\begin{itemize}

\item{The median redshift among the 38 unique sources detected in the
  full-resolution and/or 1\arcsec~tapered 10\,GHz images confirmed by
  having OIR counterparts is $\langle z \rangle = 1.24\pm0.15$. }

\item{Of the 32 sources reliably detected at 10\,GHz  
%%28 are resolved 
  in the image with $0\,\farcs22$ FWHM resolution, the weighted median of their
  deconvolved FHWM major axes is $\langle \theta_{\rm M} \rangle =
  0\,\farcs17\pm0\,\farcs03$, and the weighted rms size scatter is
  $\approx 0\,\farcs 09$.  The weighted median linear major-axis FWHM size of
  these sources is $\langle \theta_{\rm M} \rangle = 1.2\pm0.28$\,kpc, and
  the weigthed rms scatter in the linear sizes is $\approx 0.79$\,kpc.  In
  units of effective radius $r_{\rm e}$, these values are equal to
  $\langle r_{\rm e} \rangle = 69\pm 13\mathrm{\,mas\,} / 508\pm114
  \mathrm{\,pc}$ with corresponding rms scatter $\approx 38 
  \mathrm{\,mas} / 324 \mathrm{\,pc}$. We found no evidence for evolution in
  radio size with redshift.  Our 10\,GHz source sizes are
  significantly smaller than  lower-frequency radio sizes
  reported in the literature, but they appear to agree with
  high-resolution mm/sub-mm sizes that trace dust emission and
  with extinction-corrected H$\alpha$ sizes.  This result indicates that
  star formation near the cosmic star formation rate peak
  largely occurs in relatively compact regions within
  galaxies.}

\item{For the detections in our $1\arcsec$-resolution 10\,GHz image
  that have 1.4\,GHz counterparts, we measured a median spectral index
  $\langle \alpha_{\rm 1.4\,GHz}^{\rm 10\,GHz}\rangle \approx -0.74\pm0.10$,
  and the spectral indices have an rms scatter $\sigma_{\alpha_{\rm
      1.4\,GHz}^{\rm 10\,GHz}} \approx 0.35$, consistent with what is
  measured for star-forming galaxies in the local universe and the
  measurement errors in $\alpha$.  Adding the sources with only upper
  limits to their 1.4\,GHz flux densities places a lower limit on the
  median spectral index $\langle \alpha_{\rm 1.4\,GHz}^{\rm 10\,GHz}
  \rangle \gtrsim -0.61$, indicating that a signifiant fraction of
  $z\gtrsim1$ sources selected at 10\,GHz have relatively flat
  spectra, which may indicate that free-free emission contributes
  significantly and makes their total flux densities robust measures
  of the current star formation activity in such sources.  }

\item{Using the spectral indices measured for detections in the $1\arcsec$-resolution 10\,GHz image having 1.4 counterparts, and assuming a typical non-thermal spectral index for each source (i.e., $\alpha_{\rm NT} \approx -0.85$), we estimate a median thermal fraction of $26\pm0.09$\% with a standard deviation of 31\% for a median rest-frame frequency of $\approx 20$\,GHz.   
Additionally including the 8 sources having only upper limits for the 1.4\,GHz flux densities places a lower limit on the median thermal fraction of $\gtrsim$48\% at the same median rest-frame frequency of $\approx 20$\,GHz.  
}

\item{Using a combination of {\it HST}/WFC3 G140 grism and MOSFIRE
  spectroscopy, we measured a new redshift for VLA J123642$+$621331 of
  $z=2.018$ that is significantly lower than the $z=4.424$ previously reported
  in the literature.  
  Using data from the radio into into the optical, the best-fit spectral energy distribution is characterized by a stellar mass of $M_{*} = 2.4\times10^{10}\,M_{\odot}$, a stellar mass fraction of $f_{*} = 0.4$, an IR luminosity of $L_{\rm IR} = 2.3\times10^{12}\,L_{\odot}$, and an extremely hot dust temperature of $T_{\rm dust} = 70\,$K .  
}
\end{itemize}

\acknowledgements
We would like to thank the anonymous referee for very useful comments that helped to improve the content and presentation of this paper.
EJM thanks G. Brammer and P. van Dokkum for useful discussions that helped improve this paper.  
EJM also thanks K. Nyland for helping with figure preparation, D. Riechers for providing unpublished source positions to help prepare these observations, and A. Pope for providing additional data to help with the analysis.  
The National Radio Astronomy Observatory is a facility of the National Science Foundation operated under cooperative agreement by Associated Universities, Inc.
E.J.M. acknowledges the hospitality of the Aspen Center for Physics, which is supported by the National Science Foundation Grant No. PHY-1066293.  
%This research has made use of the NASA/IPAC Infrared Science Archive, which is operated by the Jet Propulsion Laboratory, California Institute of Technology, under contract with the National Aeronautics and Space Administration.
%This research has made use of the NASA/IPAC Extragalactic Database (NED), as well as the NASA/IPAC Infrared Science Archive, both of which are operated by the Jet Propulsion Laboratory, California Institute of Technology, under contract with the National Aeronautics and Space Administration.

\appendix

\section{Notes on Specific Redshifts}\label{sec:AppendixA}

The source in Table~\ref{tbl-1} with coordinates $\alpha =12^\mathrm{h}36^\mathrm{m}44\fs110, \delta =+62\degr12\arcmin44\farcs81$ (also listed in Table 2) has a grism-based redshift of $z=1.676$ reported by \citet{im16}, with a  68\% confidence interval of 1.631 to 1.705.   A deeper inspection of the 3D-HST data products suggest that this redshift estimate is primarily derived from the galaxy photometry, with little contribution from the grism data.  The extracted spectrum is truncated and covers only 6\% of the normal grism spectral range, and no obvious emission or absorption features are detected.  The corresponding photometric redshift provided by \citet{im16} is $z=1.694$, with a similar 68\% confidence interval.
%While the source in Table~\ref{tbl-1} with coordinates $\alpha =12^\mathrm{h}36^\mathrm{m}44\fs110, \delta =+62\degr12\arcmin44\farcs81$ (also listed in Table 2) has a grism-based redshift of $z=1.676$ with a  68\% confidence interval of 1.631 to 1.705 given in \citet{im16}, a deeper inspection of their data products suggest that the grism data did not contribute significantly to the redshift determination. The fraction of the spectrum within the image for this object is only 6\%, with no obvious lines detected.  The corresponding photometric redshift provided by \citet{im16} is $z=1.694$ having a similar 68\% confidence interval. 

For one source included in Table~\ref{tbl-1} with coordinates $\alpha =12^\mathrm{h}37^\mathrm{m}02\fs539, \delta =+62\degr13\arcmin02\farcs32$ 
%$12~37~02.539$~$+62~13~02.32$ 
(also listed in Table~\ref{tbl-2}), the photometric redshift reported in \citet{im16} is $z = 5.04$, which is unusually high compared to the other 10\,GHz sources.  The galaxy is extremely red (see Figure~\ref{fig:srcs}) and surprisingly bright in {\it Spitzer}/IRAC for such a high redshift.  The CANDELS team photometric redshift (D. Kodra et al.\ in preparation; G. Barro et al.\ in preparation) for this galaxy is $z = 2.65$, with a 95\% confidence interval 2.21 to 3.16, which we believe to be more reliable.  

In Table~\ref{tbl-1}, the source with coordinates $\alpha =12^\mathrm{h}36^\mathrm{m}40\fs306, \delta =+62\degr13\arcmin31\farcs14$ 
%$12~36~40.306$~$+62~13~31.14$ 
is taken to be at $z=0.484$ based on a Keck/LRIS specrum \citep[][]{jcohen00,lc04}.  \citet{bcw08} report $z=0.4352$ based on Keck/DEIMOS data, while \citet{gw04} observed this galaxy but did not measure a redshift.  A. Barger (private communication) reports that the LRIS spectrum is higher quality, but that the redshift is nevertheless uncertain.

The source in Table \ref{tbl-1} with coordinates $\alpha =12^\mathrm{h}36^\mathrm{m}57\fs375, \delta =+62\degr14\arcmin07\farcs86$ 
%$12~36~57.375$~$+62~14~07.86$ 
has a tentative (``B-grade'') $z=1.460$ redshift from unpublished Keck/DEIMOS spectrum (D. Stern et al.\ in preparation) based on [OII]~3727\AA~emission.  

In Table~\ref{tbl-1} the source with coordinates $\alpha =12^\mathrm{h}36^\mathrm{m}46\fs063, \delta =+62\degr14\arcmin48\farcs70$ 
%$12~36~46.063$~$+62~14~48.70$ 
(also listed in Table \ref{tbl-2}) has a secure (``A-grade'') $z=2.003$ redshift from unpublished Keck/LRIS spectrum (D. Stern et al.\ in preparation) based on detections of Lyman~$\alpha$ and CIII]~1909\AA.  This galaxy was additionally detected by SCUBA at 850$\,\mu$m \citep[GN12 in ][]{ap05}.  

In Table~\ref{tbl-2}, the source with coordinates $\alpha =12^\mathrm{h}36^\mathrm{m}44\fs010, \delta =+62\degr14\arcmin50\farcs77$ 
%$12~36~44.010$~$+62~14~50.77$ 
has a tentative (``B-grade'') redshift of $z=1.784$ from an unpublished Keck/LRIS spectrum (D. Stern et al.\ in preparation) using CIV~1549\AA\ and FeII~2600\AA\ absorption lines.  This value is consistent with $z = 1.77$ measured from {\it Spitzer}/IRS spectroscopy \citep{akirk12}, which shows silicate 9.7$\,\mu$m absorption indicating the presence of an obscured AGN.  \citet{nr06} also published a redshift of $z=2.095$ for this source based on Keck/LRIS spectroscopy, however it is marked as uncertain in their data table.

\section{Relating Radio and Optical Source Sizes}\label{sec:AppendixB}

The point-spread function or ``beam'' of a telescope is usually a
circular Gaussian, and radio astronomers usually describe its
resolution in terms of its FWHM beamwidth $\theta_{1/2}$ defined
by:
\begin{equation}
\exp \Biggl[ - \frac {(\theta_{1/2} / 2)^2}{2 \sigma^2} \Biggr] = 
\frac{1}{2}~,
\end{equation}
where $\sigma$ is the rms width of the Gaussian and $\sigma^2$ is its 
variance.  Thus
\begin{equation}
\theta_{1/2} = ( 8 \ln 2)^{1/2} \sigma \approx 2.35482 \sigma~.
\end{equation}
The apparent brightness distribution of a source in an image is the
convolution of its actual brightness distribution with the beam.  If
the source is only slightly resolved, the image brightness
distribution is nearly Gaussian and an elliptical Gaussian fit to the
image brightness distribution can be used to estimate the major and
minor FWHM axes of the actual ``deconvolved'' source.  Variances add
under convolution, so the deconvolved FWHM major and minor axes are
\begin{align}\label{eqn:deconv}
\theta_\mathrm{M}^2  = & \,\phi_\mathrm{M}^2 - \theta_{1/2}^2 \\
\theta_\mathrm{m}^2  = & \,\phi_\mathrm{m}^2 - \theta_{1/2}^2 ~,
\end{align}
where $\phi_\mathrm{M}$ and $\phi_\mathrm{m}$ are the image FWHM
sizes.  Equations~\ref{eqn:deconv} and A4 with $\theta_{1/2} =
0\,\farcs22$ were used to calculate the values of $\theta_\mathrm{M}$
and $\theta_\mathrm{m}$ listed in Table~\ref{tbl-1}.
For a non-Gaussian source brightness distribution, $\theta_\mathrm{M}$
and $\theta_\mathrm{m}$ may not be FWHM sizes; rather, they indicate
only the variance of the source brightness distribution:
$\sigma^2_\mathrm{M} = \theta_\mathrm{M}^{2} / (8 \ln 2)$ and
$\sigma^2_\mathrm{m} = \theta_\mathrm{m}^{2} / (8 \ln 2)$.

Most $\mu$Jy radio sources are powered by star-forming galaxies whose
face-on radio brightness distributions are better approximated by a
transparent thin circular disk whose normalized brightness declines
exponentially with some scale length $\beta$:
\begin{equation}\label{eqn:expdisk}
B(r) = \frac {1} {2 \pi \beta^2} \exp \Biggl( - \frac{r}{\beta} \Biggr)~.
\end{equation}
Optical astronomers typically specify the disk size in terms of its
effective radius $r_\mathrm{e}$, defined as the radius enclosing half
of the total flux density.
\begin{equation}\label{eqn:redef}
\frac{1}{2} \equiv 2 \pi \int_0^{r_\mathrm{e}} B(r) r dr~.
\end{equation}
The relation between $\beta$ and $r_\mathrm{e}$ can be calculated by
inserting Equation~\ref{eqn:expdisk} into Equation~\ref{eqn:redef} and
integrating to get
\begin{equation}\label{eqn:rebeta}
\frac{1}{2} = \Biggl( 1 + \frac{r_\mathrm{e}} {\beta} \Biggr)
\exp \Biggl( - \frac {r_\mathrm{e}} {\beta} \Biggr)~.
\end{equation}
Solving Equation~\ref{eqn:rebeta} numerically yields $r_\mathrm{e}
\approx 1.67835 \beta$.  As $\beta$ is the scale length over which
the brightness declines by a factor of $e^1 \approx 10^{0.434}$, 
$r_\mathrm{e}$ is the scale length over which the brightness declines
by a factor of $e^{1.67835} \approx 10^{0.729}$.

The variance $\langle x^2 \rangle$ of a circular exponential disk is
\begin{align}
\langle x^2 \rangle \equiv & \int_0^\infty  \int_{\psi=0}^{2 \pi}
x^2 B(r) r \, d \psi \, dr = 
\int_{r = 0}^\infty \int_{\psi - 0}^{2 \pi} 
[r \cos(\psi)]^2 B(r) r \, d \psi \, dr \\
\langle x^2 \rangle  = & \frac{1}{2 \pi \beta^2} \int_{\psi = 0}^{2 \pi}
\cos^2(\psi) \, d \psi \int_{r = 0}^\infty r^3 \exp \Biggl( -
\frac{r}{\beta} \Biggr) \, dr = 
\frac{1}{2 \beta^2} \int_0^\infty r^3 \exp \Biggl( - 
\frac {r}{\beta} \Biggr) \, dr~.
\end{align}
Integrating by parts three times gives
\begin{equation}
\langle x^2 \rangle = 3 \beta^2 \approx 1.0650 r_\mathrm{e}^2
~.
\end{equation}

If a transparent thin face-on circular exponential disk is tilted by
inclination angle $i$, it appears as an elliptical exponential disk
whose unchanged major axis $\theta_\mathrm{M}$ measures the disk
$r_\mathrm{e}$ and whose minor axis $\theta_\mathrm{m}$ has been
shortened by the factor $\cos(i)$.  In the limit $r_\mathrm{e} <
\theta_{1/2}$, the image of a tilted exponential disk galaxy is a
nearly Gaussian ellipse, and the deconvolved disk major axis is
related to the effective radius by
\begin{equation}\label{eqn:fwhmre}
\theta_\mathrm{M} = ( 8 \ln 2 \,\langle x^2 \rangle )^{1/2}
\approx 2.430 r_\mathrm{e}~.
\end{equation}

\section{The Ratio of $S_\mathrm{P} / S_\mathrm{I}$ for an Elliptical
Exponential Disk Observed with a Circular Gaussian Beam}\label{sec:AppendixC}

If a circular Gaussian beam with attenuation power pattern
\begin{equation}
A(r) = \exp \Biggl( - \frac {r^2}{2 \sigma^2} \Biggr)
\end{equation}
is pointed at a circular source with brightness distribution $B(r)$,
the attenuated peak brightness on the image is 
\begin{equation}
S_\mathrm{a} = 
2 \pi \int_0^\infty A(r) B(r) \,r \, dr~.
\end{equation}
Inserting the brightness distribution of a circular exponential
disk (Equation~\ref{eqn:expdisk}) gives the ratio of
peak brightness $S_\mathrm{P}$ to the integrated flux density
$S_\mathrm{I}$:
\begin{equation}
\frac{S_\mathrm{P}}{S_\mathrm{I}} = S_\mathrm{a} = 
\frac {2 \pi}{2 \pi \beta^2} \int_0^\infty
\exp \Biggl( - \frac{r^2}{2\sigma^2} \Biggr)
\exp \biggl( - \frac{r}{\beta} \Biggr) \,r\,dr =
\frac{1}{\beta^2} \int_0^\infty r
\exp\Biggl[ - \frac{1}{2\sigma^2} \Biggl( r^2 + 
\frac{2\sigma^2}{\beta} r \Biggr) \Biggr]\,dr ~.
\end{equation}
This integral can be evaluated in terms of the complementary error
function
\begin{equation}
\mathrm{erfc}(z) \equiv 1 - \mathrm{erf}(z) = \frac {2}{\sqrt{\pi}}
\int_z^\infty \exp(-t^2) \,dt
\end{equation}
for
\begin{equation}
z = s^{-1/2} \Biggl(\frac{\sigma}{\beta} \Biggr) \approx
0.50398 \Biggl( \frac{\theta_{1/2}}{r_\mathrm{e}} \Biggr)~.
\end{equation}
The result is
\begin{equation}\label{eqn:spsi}
\frac{S_\mathrm{P}} {S_\mathrm{I}} = 
2 z^2 [ 1 - \sqrt{\pi} \,z \, \exp(z^2) \mathrm{erfc}(z)]~.
\end{equation}
Equation~\ref{eqn:spsi} is exact only for a circular exponential disk.
For an elliptical exponential disk, a very good approximation is the geometric
mean of the $S_\mathrm{P} / S_\mathrm{I}$ values calculated for circular
disks matching the $z$ values calculated for the major (M) and minor (m)
axes of the ellipse:
\begin{equation}\label{eqn:spsiapprox}
\frac {S_\mathrm{P}}{S_\mathrm{I}} \approx \Biggl[ 
\biggl( \frac{S_\mathrm{P}}{S_\mathrm{I}} \biggr)_\mathrm{M}
\biggl( \frac{S_\mathrm{P}}{S_\mathrm{I}} \biggr)_\mathrm{m}\Biggr]^{1/2}~.
\end{equation}

\newpage

%\bibliography{/Users/emurphy/libs/bibtexref/master_ref}
\bibliography{aph.bbl}

%\LongTables
\clearpage
\begin{turnpage}
\input{tab2}
\end{turnpage}

\end{document}

%% file: tab1.tex
\begin{deluxetable*}{cc|lcccrrrc}
\tablecaption{Full Resolution 10\,GHz Source Characteristics \label{tbl-1}}
\tabletypesize{\scriptsize}
\tablewidth{0pt}
\tablehead{
\colhead{R.A.} & \colhead{Decl.} & \colhead{$z$} & \colhead{$z^{a}$} &\colhead{$r^{b}$} & \colhead{$JH_{\rm NIR}~^{c}$} &
\colhead{~$S_{\rm P}$} & \colhead{~~$S_{\rm I}$} & \colhead{~~$S_{*}$} & \colhead{$\theta_{\rm M}\times\theta_{\rm m}$}\\
\colhead{(J2000)} & \colhead{(J2000)} & \colhead{} & \colhead{\llap{(}type\rlap{)}} & \colhead{(mas)} & \colhead{(mag)} &
\colhead{($\mu$Jy\,beam$^{-1}$\rlap{)}} & \colhead{~~($\mu$Jy)} & \colhead{~~($\mu$Jy)} & \colhead{(mas)}
}
\multicolumn{10}{c}{$S_{\rm P}/\sigma_{\rm n} \geq 5$}\\[2pt]
\hline
\\[-6pt]
                  $12~36~34.211$   &              $+62~14~32.95$  &          0.5184  &   1  &    31  &19.52  & $  7.72\pm  1.24$  & $  8.71\pm  2.09$  &           $  8.20\pm  1.24$  &      $148\pm86\times 0\pm118$\\
 \llap{$^{\rm d}$}$12~36~34.514$   &              $+62~12~41.08$  &          1.2234  &   1  &   139  &21.11  & $  9.02\pm  1.23$  & $ 27.90\pm  4.87$  &   $ 27.90\pm  4.87$\rlap{$^{\rm e}$}  &     $354\pm69\times 283\pm59$\\
                  $12~36~42.091$   &              $+62~13~31.43$  &          2.018 &   1  &    16  &23.76  & $ 37.81\pm  0.70$  & $ 36.80\pm  1.22$  &           $ 37.30\pm  0.70$  &        $42\pm23\times 0\pm41$\\
                  $12~36~42.214$   &              $+62~15~45.51$  &          0.8575  &   1  &   173  &20.02  & $ 27.78\pm  1.47$  & $ 32.87\pm  2.42$  &   $ 32.87\pm  2.42$\rlap{$^{\rm e}$}  &       $144\pm28\times 0\pm69$\\
                  $12~36~44.110$   &              $+62~12~44.81$  &          1.676  &   2  &    49  &21.85  & $  3.97\pm  0.74$  & $  4.46\pm  1.25$  &           $  4.21\pm  0.74$  &     $171\pm100\times 0\pm120$\\
 \llap{$^{\rm d}$}$12~36~44.386$   &              $+62~11~33.14$  &          1.0128  &   1  &    30  &19.46  & $279.25\pm  2.14$  & $340.57\pm  4.25$  &   $340.57\pm  4.25$\rlap{$^{\rm e}$}  &        $139\pm4\times 56\pm7$\\
                  $12~36~46.063$   &              $+62~14~48.70$  &          2.003  &   1  &   155  &23.71  & $  6.95\pm  0.82$  & $ 10.65\pm  1.25$  &   $ 10.65\pm  1.25$\rlap{$^{\rm e}$}  &     $214\pm57\times 101\pm62$\\
                  $12~36~46.332$   &              $+62~14~04.69$  &          0.9605  &   1  &    82  &20.11  & $ 92.69\pm  0.64$  & $ 91.06\pm  1.12$  &   $ 91.06\pm  1.12$\rlap{$^{\rm e}$}  &         $53\pm7\times 0\pm25$\\
                  $12~36~48.076$   &              $+62~13~09.01$  &          0.4745  &   1  &    35  &19.59  & $  4.22\pm  0.64$  & $  4.38\pm  1.09$  &           $  4.30\pm  0.64$  &      $80\pm108\times 0\pm121$\\
                  $12~36~48.330$   &              $+62~14~16.57$  &          2.002  &   1  &   149  &22.55  & $  3.20\pm  0.63$  & $  3.15\pm  1.13$  &           $  3.17\pm  0.63$  &     $146\pm114\times 0\pm118$\\
                  $12~36~52.884$   &              $+62~14~44.07$  &          0.3208  &   1  &    18  &18.22  & $114.69\pm  0.72$  & $113.24\pm  1.25$  &           $113.96\pm  0.72$  &        $25\pm13\times 0\pm24$\\
                  $12~36~53.367$   &              $+62~11~39.58$  &          1.268  &   1  &    45  &21.47  & $  5.43\pm  1.03$  & $ 20.82\pm  1.28$  &   $ 20.82\pm  1.28$\rlap{$^{\rm e}$}  &    $553\pm140\times 220\pm73$\\
                  $12~36~55.449$   &              $+62~13~11.24$  &          0.9544  &   1  &    11  &20.63  & $ 20.11\pm  0.64$  & $ 20.53\pm  1.11$  &           $ 20.32\pm  0.64$  &        $54\pm31\times 0\pm55$\\
                  $12~36~56.914$   &              $+62~13~01.64$  &          1.2409  &   1  &    21  &21.16  & $  8.69\pm  0.66$  & $  8.12\pm  1.17$  &           $  8.40\pm  0.66$  &       $24\pm160\times 0\pm82$\\
                  $12~36~58.843$   &              $+62~14~34.92$  &          0.6766  &   1  &    34  &19.75  & $  4.71\pm  0.77$  & $  4.86\pm  1.32$  &           $  4.78\pm  0.77$  &      $110\pm98\times 0\pm119$\\
                  $12~37~04.873$   &              $+62~16~01.55$  &          1.170  &   1  &    16  &23.20  & $ 12.83\pm  2.24$  & $ 13.02\pm  3.93$  &           $ 12.92\pm  2.24$  &      $145\pm99\times 0\pm113$\\
                  $12~37~11.251$   &              $+62~13~30.87$  &          1.9958  &   1  &    11  &22.40  & $  7.48\pm  1.35$  & $  8.15\pm  2.34$  &           $  7.81\pm  1.35$  &     $181\pm100\times 0\pm113$\\
                  $12~37~11.984$   &              $+62~13~25.69$  &          1.992  &   1  &    92  &23.26  & $  7.30\pm  1.41$  & $  7.03\pm  2.56$  &           $  7.16\pm  1.41$  &     $152\pm113\times 0\pm113$\\
                  $12~37~16.376$   &              $+62~15~12.35$  &          0.5577  &   1  &    60  &19.15  & $ 87.07\pm  4.03$  & $ 95.58\pm  6.78$  &           $ 91.22\pm  4.03$  &       $96\pm29\times 23\pm94$\\
\hline\\[-6pt]
\multicolumn{10}{c}{$3.5 \leq S_{\rm P}/\sigma_{\rm n} < 5$}\\[2pt]
\hline
\\[-6pt]
                  $12~36~19.565$   &              $+62~13~42.93$  &          1.699  &   3  &   159  &25.43  & $ 31.16\pm  6.76$  & $ 31.68\pm 11.85$  &           $ 31.42\pm  6.76$  &     $141\pm124\times 0\pm128$\\
                  $12~36~27.872$   &              $+62~14~49.08$  &          0.6802  &   1  &    50  &19.96  & $ 12.78\pm  2.65$  & $ 25.68\pm  3.78$  &           $ 18.12\pm  2.65$  &    $318\pm112\times 122\pm93$\\
                  $12~36~35.592$   &              $+62~14~24.04$  &          2.0150  &   1  &    64  &21.45  & $  4.90\pm  1.12$  & $ 12.32\pm  1.51$  &           $  7.77\pm  1.12$  &    $377\pm131\times 171\pm91$\\
                  $12~36~40.306$   &              $+62~13~31.14$  &          0.484  &   1  &    66  &21.24  & $  3.39\pm  0.72$  & $  3.88\pm  1.26$  &           $  3.63\pm  0.72$  &     $238\pm126\times 0\pm112$\\
                  $12~36~41.604$   &              $+62~13~49.41$  &          3.244  &   3  &    62  &24.37  & $  3.20\pm  0.76$  & $  5.01\pm  1.16$  &           $  4.00\pm  0.76$  &    $246\pm121\times 65\pm166$\\
                  $12~36~42.128$   &              $+62~13~48.34$  &          0.817  &   3  &    81  &25.32  & $  3.25\pm  0.71$  & $  3.41\pm  1.22$  &           $  3.33\pm  0.71$  &     $119\pm127\times 0\pm138$\\
                  $12~36~46.736$   &              $+62~14~45.84$  &          2.004  &   1  &    93  &22.47  & $  3.45\pm  0.75$  & $  3.58\pm  1.29$  &           $  3.51\pm  0.75$  &      $87\pm148\times 0\pm146$\\
                  $12~36~48.524$   &              $+62~14~36.91$  &          1.365  &   3  &    88  &26.68  & $  2.93\pm  0.69$  & $  2.95\pm  1.21$  &           $  2.94\pm  0.69$  &     $126\pm138\times 0\pm138$\\
                  $12~36~49.688$   &              $+62~13~13.01$  &          0.4745  &   1  &   179  &20.61  & $  2.44\pm  0.66$  & $  3.32\pm  1.03$  &           $  2.84\pm  0.66$  &    $173\pm131\times 85\pm164$\\
                  $12~36~57.375$   &              $+62~14~07.86$  &          1.460  &   1  &    73  &22.03  & $  2.95\pm  0.65$  & $  3.21\pm  1.14$  &           $  3.07\pm  0.65$  &     $199\pm125\times 0\pm120$\\
                  $12~36~59.614$   &              $+62~11~53.36$  &          1.0205  &   1  &    55  &21.76  & $  4.25\pm  1.11$  & $  6.00\pm  1.75$  &           $  5.05\pm  1.11$  &     $237\pm137\times 0\pm150$\\
                  $12~37~02.539$   &              $+62~13~02.32$  &          2.650  &   4  &   145  &25.63  & $  4.14\pm  0.83$  & $  6.84\pm  1.26$  &           $  5.32\pm  0.83$  &     $292\pm112\times 0\pm131$\\
                  $12~37~08.748$   &              $+62~12~57.83$  &          2.268  &   1  &   107  &23.12  & $  5.89\pm  1.22$  & $  7.04\pm  2.01$  &           $  6.44\pm  1.22$  &     $162\pm107\times 0\pm137$
\enddata
\tablenotetext{a}{Redshift type:  (1) Spectroscopic (\citealp{jcohen00,gw04, ams04, tt05, nr06, bcw08,dtf08, ht11}; D. Stern et al.\ in preparation; this paper); (2) Grism-based \citep{im16}; (3) Photometric \citep{im16}; (4) Photometric (D. Kodra et al.\ in preparation; G. Barro et al.\ in preparation)}
\tablenotetext{b}{The angular separation between the 10\,GHz detection and the OIR counterpart.}
\tablenotetext{c}{NIR magnitude from some combination of $J_{125}$, $JH_{140}$ and $H_{160}$ {\it HST}/WFC3 images scaled to the $JH_{140}$ AB zeropoint as described in \citep{im16}.}
\tablenotetext{d}{Single-Gaussian, fitted parameters reported by {\sc imfit} since the {\sc PyBDSM} fit included multiple Gaussian components.}
\tablenotetext{e}{Confidently ($\geq2\sigma_{\phi}$) resolved.}
\end{deluxetable*}

%% file: tab2.tex
\begin{deluxetable*}{cc|lcccrrrccc}
\tablecaption{1\arcsec~Tapered 10\,GHz Source Characteristics \label{tbl-2}}
\tabletypesize{\scriptsize}
\tablewidth{0pt}
\tablehead{
\colhead{R.A.} & \colhead{Decl.} & \colhead{$z$} & \colhead{$z^{a}$} & \colhead{$r^{b}$} & \colhead{\llap{$J$}$H_{\rm NIR}$~\rlap{$^{c}$}} &
\colhead{$S_{\rm P}$} & \colhead{~~$S_{\rm I}$} & \colhead{~~$S_{*}$} & \colhead{$\theta_{\rm M}\times\theta_{\rm m}$} & \colhead{$\alpha^{\rm 10\,GHz}_{\rm 1.4\,GHz}$} & \colhead{$f_{\rm T}^{\nu_{\rm r},{\rm d}}$}\\
\colhead{(J2000)} & \colhead{(J2000)} & \colhead{} & \colhead{\llap{(ty}p\rlap{e)}} &  \colhead{} & \colhead{(mag)} &
\colhead{\llap{(}$\mu$Jy\,beam$^{-1}$\rlap{)}} & \colhead{~~($\mu$Jy)} & \colhead{~~($\mu$Jy)} & \colhead{} & \colhead{} & \colhead{}
}
\multicolumn{12}{c}{$S_{\rm P}/\sigma_{\rm n} \geq 5$}\\[2pt]
\hline
\\[-6pt]
 \llap{$^{\rm e}$}$12~36~34.460$   &       $\llap{+}62~12~12.93$  &          0.4573  &  \llap{1}  & 0\farcs16  &18.12  & $ 22.66\pm  2.97$  & $ 49.58\pm  4.13$  &    $ 49.58\pm  4.13$\rlap{$^{\rm f}$}  &         $1\farcs37\pm0\farcs30\times 0\farcs82\pm0\farcs25$  & $ -0.77\pm  0.05$  & $  0.19\pm  0.11$\\
                  $12~36~34.502$   &       $\llap{+}62~12~41.04$  &          1.2234  &  \llap{1}  & 0\farcs19  &21.11  & $ 27.03\pm  2.42$  & $ 36.71\pm  3.80$  &                     $ 31.50\pm  2.42$  &         $0\farcs68\pm0\farcs20\times 0\farcs51\pm0\farcs21$  & $ -0.94\pm  0.05$  & $  0.21\pm  0.09$\\
                  $12~36~35.594$   &       $\llap{+}62~14~24.16$  &          2.011  &  \llap{1}  & 0\farcs18  &21.45  & $ 12.07\pm  2.12$  & $ 18.16\pm  3.27$  &                     $ 14.80\pm  2.12$  &         $1\farcs03\pm0\farcs40\times 0\farcs32\pm0\farcs53$  & $ -0.83\pm  0.09$  & $  0.04\pm  0.22$\\
                  $12~36~42.084$   &       $\llap{+}62~13~31.41$  &          2.018  &  \llap{1}  & 0\farcs03  &23.76  & $ 52.37\pm  1.37$  & $ 67.01\pm  2.20$  &    $ 67.01\pm  2.20$\rlap{$^{\rm f}$}  &         $0\farcs65\pm0\farcs06\times 0\farcs39\pm0\farcs07$  & $ -1.02\pm  0.02$  & $  0.21\pm  0.04$\\
                  $12~36~42.215$   &       $\llap{+}62~15~45.50$  &          0.8575  &  \llap{1}  & 0\farcs18  &20.02  & $ 36.34\pm  2.77$  & $ 41.84\pm  4.60$  &                     $ 38.99\pm  2.77$  &         $0\farcs58\pm0\farcs19\times 0\farcs00\pm0\farcs38$  & $ -0.69\pm  0.05$  & $  0.35\pm  0.09$\\
 \llap{$^{\rm g}$}$12~36~44.385$   &       $\llap{+}62~11~33.13$  &          1.0128  &  \llap{1}  & 0\farcs04  &19.46  & $426.08\pm  3.42$  & $449.83\pm  5.83$  &    $449.83\pm  5.83$\rlap{$^{\rm f}$}  &         $0\farcs69\pm0\farcs05\times 0\farcs00\pm0\farcs25$  & $ -0.71\pm  0.02$  & $  0.32\pm  0.03$\\
 \llap{$^{\rm g}$}$12~36~46.052$   &       $\llap{+}62~14~48.73$  &          2.003  &  \llap{1}  & 0\farcs16  &23.71  & $ 13.33\pm  2.19$  & $ 35.13\pm  2.90$  &    $ 35.13\pm  2.90$\rlap{$^{\rm f}$}  &         $2\farcs75\pm0\farcs71\times 2\farcs36\pm0\farcs65$  & $ -0.58\pm  0.05$  & $  0.54\pm  0.07$\\
 \llap{$^{\rm g}$}$12~36~46.332$   &       $\llap{+}62~14~04.70$  &          0.9605  &  \llap{1}  & 0\farcs08  &20.11  & $125.72\pm  1.74$  & $138.89\pm  2.91$  &    $138.89\pm  2.91$\rlap{$^{\rm f}$}  &         $0\farcs75\pm0\farcs09\times 0\farcs53\pm0\farcs11$  & $ -0.43\pm  0.02$  & $  0.73\pm  0.02$\\
                  $12~36~49.692$   &       $\llap{+}62~13~13.03$  &          0.4745  &  \llap{1}  & 0\farcs15  &20.61  & $  6.61\pm  1.20$  & $ 13.30\pm  1.71$  &    $ 13.30\pm  1.71$\rlap{$^{\rm f}$}  &         $1\farcs46\pm0\farcs45\times 0\farcs54\pm0\farcs37$  & $ -0.76\pm  0.10$  & $  0.21\pm  0.21$\\
 \llap{$^{\rm h}$}$12~36~52.876$   &       $\llap{+}62~14~44.06$  &          0.3208  &  \llap{1}  & 0\farcs05  &18.22  & $137.92\pm  1.59$  & $191.51\pm  3.46$  &    $191.51\pm  3.46$\rlap{$^{\rm f}$}  &         $0\farcs66\pm0\farcs03\times 0\farcs59\pm0\farcs03$  & $ -0.02\pm  0.03$  & $  1.00\pm  0.01$\\
                  $12~36~53.361$   &       $\llap{+}62~11~39.57$  &          1.268  &  \llap{1}  & 0\farcs05  &21.47  & $ 16.00\pm  2.07$  & $ 19.02\pm  3.41$  &                     $ 17.45\pm  2.07$  &         $0\farcs61\pm0\farcs31\times 0\farcs16\pm0\farcs77$  & $ -0.80\pm  0.08$  & $  0.12\pm  0.19$\\
                  $12~36~55.448$   &       $\llap{+}62~13~11.21$  &          0.9544  &  \llap{1}  & 0\farcs02  &20.63  & $ 18.30\pm  1.13$  & $ 15.93\pm  2.05$  &                     $ 17.07\pm  1.13$  &         $0\farcs00\pm0\farcs36\times 0\farcs00\pm0\farcs33$  &      $\geq -0.07$  &      $\geq  1.00$\\
                  $12~36~56.919$   &       $\llap{+}62~13~01.76$  &          1.2409  &  \llap{1}  & 0\farcs10  &21.16  & $  8.52\pm  1.28$  & $ 13.93\pm  1.95$  &    $ 13.93\pm  1.95$\rlap{$^{\rm f}$}  &         $1\farcs29\pm0\farcs37\times 0\farcs07\pm1\farcs76$  &      $\geq -0.17$  &      $\geq  0.96$\\
                  $12~37~16.372$   &       $\llap{+}62~15~12.32$  &          0.5577  &  \llap{1}  & 0\farcs03  &19.15  & $113.95\pm  7.81$  & $132.98\pm 12.92$  &    $132.98\pm 12.92$\rlap{$^{\rm f}$}  &         $0\farcs64\pm0\farcs16\times 0\farcs00\pm0\farcs36$  & $ -0.05\pm  0.06$  & $  1.00\pm  0.03$\\
\hline\\[-6pt]
\multicolumn{12}{c}{$3.5 \leq S_{\rm P}/\sigma_{\rm n} < 5$}\\[2pt]
\hline
\\[-6pt]
                  $12~36~27.861$   &       $\llap{+}62~14~49.07$  &          0.6802  &    \llap{1}  & 0\farcs13  &19.96  & $ 18.92\pm  4.73$  & $ 18.72\pm  8.40$  &                     $ 18.82\pm  4.73$  &         $0\farcs64\pm0\farcs66\times 0\farcs00\pm0\farcs62$  &      $\geq -0.02$  &      $\geq  1.00$\\
                  $12~36~34.227$   &       $\llap{+}62~14~33.09$  &          0.5184  &    \llap{1}  & 0\farcs18  &19.52  & $  9.14\pm  2.19$  & $  8.40\pm  4.04$  &                     $  8.77\pm  2.19$  &         $0\farcs64\pm0\farcs66\times 0\farcs00\pm0\farcs56$  & $ -0.54\pm  0.15$  & $  0.59\pm  0.21$\\
 \llap{$^{\rm e}$}$12~36~43.963$   &       $\llap{+}62~12~50.08$  &          0.557  &    \llap{1}  & 0\farcs07  &20.13  & $  5.01\pm  1.35$  & $  8.04\pm  2.09$  &                     $  6.35\pm  1.35$  &         $1\farcs43\pm0\farcs73\times 0\farcs00\pm0\farcs65$  & $ -0.89\pm  0.16$  & $  0.22\pm  0.31$\\
 \llap{$^{\rm e}$}$12~36~44.010$   &       $\llap{+}62~14~50.77$  &          1.784  &    \llap{1}  & 0\farcs23  &22.62  & $  6.25\pm  1.60$  & $  9.33\pm  2.48$  &                     $  7.63\pm  1.60$  &         $1\farcs10\pm0\farcs60\times 0\farcs09\pm2\farcs57$  & $ -0.74\pm  0.16$  & $  0.26\pm  0.32$\\
                  $12~36~44.101$   &       $\llap{+}62~12~44.58$  &          1.676  &    \llap{2}  & 0\farcs26  &21.85  & $  5.79\pm  1.39$  & $  8.13\pm  2.22$  &                     $  6.86\pm  1.39$  &         $1\farcs16\pm0\farcs60\times 0\farcs00\pm0\farcs62$  &      $\geq -0.53$  &      $\geq  0.60$\\
 \llap{$^{\rm e}$}$12~36~46.377$   &       $\llap{+}62~16~29.56$  &          0.5032  &    \llap{1}  & 0\farcs33  &19.36  & $ 22.99\pm  4.81$  & $ 29.50\pm  7.71$  &                     $ 26.04\pm  4.81$  &         $0\farcs70\pm0\farcs48\times 0\farcs33\pm0\farcs66$  & $ -1.43\pm  0.10$  & $  0.19\pm  0.16$\\
                  $12~36~48.292$   &       $\llap{+}62~14~16.59$  &          2.002  &    \llap{1}  & 0\farcs10  &22.55  & $  4.55\pm  1.26$  & $  6.32\pm  1.98$  &                     $  5.37\pm  1.26$  &         $0\farcs81\pm0\farcs61\times 0\farcs40\pm0\farcs75$  &      $\geq -0.66$  &      $\geq  0.41$\\
 \llap{$^{\rm e}$}$12~36~50.104$   &       $\llap{+}62~14~01.08$  &          2.231  &    \llap{1}  & 0\farcs07  &23.65  & $  5.57\pm  1.20$  & $  7.54\pm  1.89$  &                     $  6.48\pm  1.20$  &         $0\farcs62\pm0\farcs49\times 0\farcs57\pm0\farcs50$  &      $\geq -0.56$  &      $\geq  0.56$\\
 \llap{$^{\rm e}$}$12~36~55.026$   &       $\llap{+}62~12~52.22$  &          0.9929  &    \llap{3}  & 0\farcs12  &25.09  & $  4.71\pm  1.30$  & $  7.26\pm  1.99$  &                     $  5.85\pm  1.30$  &         $1\farcs04\pm0\farcs62\times 0\farcs37\pm0\farcs75$  &      $\geq -0.61$  &      $\geq  0.48$\\
                  $12~36~58.844$   &       $\llap{+}62~14~34.97$  &          0.6766  &    \llap{1}  & 0\farcs09  &19.75  & $  6.43\pm  1.47$  & $  8.08\pm  2.38$  &                     $  7.21\pm  1.47$  &         $0\farcs76\pm0\farcs53\times 0\farcs00\pm0\farcs67$  &      $\geq -0.51$  &      $\geq  0.64$\\
                  $12~37~02.547$   &       $\llap{+}62~13~02.18$  &          2.650  &    \llap{4}  & 0\farcs17  &25.63  & $  5.89\pm  1.53$  & $ 13.52\pm  2.13$  &                     $  8.92\pm  1.53$  &         $1\farcs82\pm0\farcs75\times 0\farcs48\pm0\farcs55$  & $ -0.54\pm  0.12$  & $  0.60\pm  0.17$\\
                  $12~37~04.894$   &       $\llap{+}62~16~01.46$  &          1.170  &    \llap{1}  & 0\farcs18  &23.20  & $ 19.30\pm  4.39$  & $ 19.92\pm  7.52$  &                     $ 19.61\pm  4.39$  &         $0\farcs27\pm0\farcs93\times 0\farcs00\pm0\farcs70$  & $ -0.19\pm  0.14$  & $  0.94\pm  0.10$\\
                  $12~37~11.321$   &       $\llap{+}62~13~30.86$  &          1.9958  &    \llap{1}  & 0\farcs22  &21.70  & $ 11.70\pm  2.63$  & $ 22.64\pm  3.86$  &                     $ 16.28\pm  2.63$  &         $1\farcs67\pm0\farcs64\times 0\farcs00\pm0\farcs62$  & $ -1.06\pm  0.09$  & $  0.20\pm  0.17$
\enddata
\tablenotetext{a}{Redshift type:  (1) Spectroscopic (\citealp{jcohen00,gw04, ams04, tt05, nr06, bcw08,dtf08, ht11}; D. Stern et al.\ in preparation; this paper); (2) Grism-based \citep{im16}; (3) Photometric \citep{im16}; (4) Photometric (D. Kodra et al.\ in preparation; G. Barro et al.\ in preparation)}
 \tablenotetext{b}{The angular separation between the 10\,GHz detection and the OIR counterpart.}
 \tablenotetext{c}{NIR magnitude from some combination of $J_{125}$, $JH_{140}$ and $H_{160}$ {\it HST}/WFC3 images scaled to the $JH_{140}$ AB zeropoint as described in \citep{im16}.}
 \tablenotetext{d}{Thermal fraction at the rest-frame frequency $(\nu_{\rm r}/{\rm GHz}) = 10(1+z)$.}
 \tablenotetext{e}{Not detected in the full-resolution image.}
 \tablenotetext{f}{Confidently ($\geq2\sigma_{\phi}$) resolved.}
 \tablenotetext{g}{2\arcsec~tapered image used in photometry since significantly (i.e., $>3\sigma$) more flux is recovered.}
 \tablenotetext{h}{Single-Gaussian, fitted parameters reported by {\sc imfit} since the {\sc PyBDSM} fit included multiple Gaussian components.}
\end{deluxetable*}